\documentclass[]{aa}  

\usepackage{graphicx}
\usepackage{natbib}
\usepackage[hidelinks]{hyperref}
\hypersetup{
   colorlinks = true,
   citecolor ={cyan}
}
\usepackage{txfonts}
\usepackage{color}
\usepackage{xcolor}
\usepackage[scientific-notation=true]{siunitx}
\usepackage{etoolbox}
\defcitealias{KolbRitter1990}{KR90}
\defcitealias{Ritter1988}{R88}

\begin{document} 

\authorrunning{Marchant et al.}

   \title{The role of mass transfer and common envelope evolution in the formation of merging binary black holes}

   \author{Pablo~Marchant
          \inst{1}
          \and
          Kaliro\"e~M.~W.~Pappas
          \inst{2,3}
          \and
          Monica~Gallegos-Garcia
          \inst{4,2}
          \and
          Christopher~P.~L.~Berry
          \inst{4,2,5}
          \and
          Ronald~E.~Taam
          \inst{4,2}
          \and
          Vicky~Kalogera
          \inst{4,2}
          \and
          Philipp~Podsiadlowski
          \inst{6,7}
          }

   \institute{Institute of Astrophysics, KU Leuven, Celestijnenlaan 200D, 3001, Leuven, Belgium\\
              \email{pablo.marchant@kuleuven.be}
         \and
         Center for Interdisciplinary Exploration and Research in Astrophysics (CIERA), 1800 Sherman Ave, Evanston, IL 60201, USA
         \and
         Laboratory of Nuclear Science, Massachusetts Institute of Technology, Cambridge, MA 02139
         \and
         Department of Physics and Astronomy, Northwestern University, 2145 Sheridan Road, Evanston, IL 60208, USA
         \and
         SUPA, School of Physics and Astronomy, University of Glasgow, Glasgow G12 8QQ, UK
         \and
         Department of Physics, University of Oxford, Keble Rd, Oxford OX1 3RH, United Kingdom
         \and
         Argelander-Institut für Astronomie der Universität Bonn, Auf dem Hügel 71, D-53121 Bonn, Germany
             }

  \abstract{
     As the number of merging binary black holes observed with ground-based
     gravitational-wave detectors grows, increasingly accurate theoretical
     models are required to compare to the observed sample and disentangle
     contributions from multiple channels. In formation models
     involving isolated binary stars, important uncertainties remain regarding
     the stability of mass transfer and common-envelope evolution.
     To study some of these uncertainties, we have computed binary simulations
     using the \texttt{MESA} code consisting of a $30M_\odot$ star in a low
     metallicity ($Z_\odot/10$) environment with a black-hole companion.
     We have developed an updated prescription to compute mass transfer
     rates including the possibility of outflows from outer Lagrangian points,
     as well as a method to self-consistently determine the core--envelope
     boundary in cases where there is common-envelope evolution. We find that binaries
     survive common-envelope evolution only if unstable mass transfer happens after the formation
     of a deep convective envelope, resulting in a narrow range ($0.2~\mathrm{dex}$) in period
     for successful envelope ejection. All cases where
     binary interaction is initiated with a radiative envelope have large
     binding energies ($\sim 10^{50}~\mathrm{erg}$), and result in mergers
     during the common-envelope
     phase even under the assumption that all the internal and recombination
     energy of the envelope, as well as the energy from an inspiral, is used to
     eject the envelope. This is independent of
     whether or not helium is ignited in the core of the donor, conditions under
     which various rapid-population synthesis calculations assume a successful
     envelope ejection is possible. Moreover, we find that the critical mass
     ratio for instability is such that across a large range in initial orbital
     periods ($\sim 1$--$1000$~days) merging binary black holes can be formed via stable
     mass transfer. {A large fraction of these systems
     undergo overflow of their L$_2$ equipotential, in which case we find stable
     mass transfer produces merging binary black holes even under extreme
     assumptions of mass and angular momentum outflows}. Our conclusions are limited to the study of one donor mass at a single
     metallicity, but suggest that population synthesis calculations
     overestimate the formation rate of merging binary black holes produced by
     common-envelope
     evolution, and that stable mass transfer could dominate the formation rate
     from isolated binaries. This is in agreement with a few other recent
     studies. Further work is required to extend these results to
     different masses and metallicities, and to understand how they can be
     incorporated into rapid population synthesis calculations.
  }

   \keywords{(Stars:) binaries (including multiple): close -- Stars: massive --
   Stars: black holes -- Gravitational waves -- X-rays: binaries
               }

   \maketitle
%

\section{Introduction} \label{sec:intro}
In the last five years the discoveries from ground-based gravitational wave (GW)
detectors have driven a large effort to understand the origin of compact binary
black holes (BHs). With $50$ detections with a high significance
\citep{GWTC1,GWTC2}, the majority of the observed GW sources
correspond to binary BH mergers covering a large range
of masses $\sim5$--$100M_\odot$. Important discoveries contained in the first
Gravitational Wave Transient Catalog \citep{GWTC1} include the discovery of BHs
more massive than those detected through electromagnetic observations in the
Galaxy \citep{Abbott_GW150914_2016} and the direct association of short gamma-ray
bursts with binary
neutron star mergers \citep{GW170817}.
The increased size of the sample on the second Gravitational Wave Transient Catalog \citep{GWTC2} points out to additional features in the properties
of merging binary BHs, indicative of the contribution of multiple formation
channels \citep{LIGOpop2,Zevin+2020}. Accurate predictions from different formation channels are then
necessary to understand their relative contributions.

A large number of formation scenarios have been proposed to form the merging
binary BHs observed by ground-based detectors. Scenarios that involve binary systems
include evolution through a common-envelope (CE) phase
\citep[e.g.,][]{Paczynski1976,vandenHeuvel1976,TutukovYungelson1993,
Belczynski+2002,Dominik+2012,Stevenson+2017,GiacobboMapelli2018},
chemically homogeneous evolution
\citep{MandeldeMink2016,Marchant+2016,deMinkMandel2016,duBuisson+2020,Riley+2020}, stable mass transfer
\citep[MT;][]{vandenHeuvel+2017,Neijssel+2019, Bavera+2020b} and Population III stars
\citep{Belczynski+2004, Kinugawa+2014,Inayoshi+2017}. Dynamical processes are
also predicted to contribute to the observed sample, including isolated triple
systems \citep{Thompson2011,Antonini+2017,myclone+2020} and interactions in globular
\citep[e.g.,
][]{Kulkarni+1993,SigurdssonHernquist1993,PortegieszwartMcmillan2000,Rodriguez+2015,diCarlo+2019} and
nuclear clusters \citep{AntoniniPerets2012}. Additional formation scenarios include
the pairing and growth of stellar mass BHs in the disks of active galactic
nuclei \citep{mcKernan+2014,Stone+2017,Bartos+2017}, and primordial BH formation
\citep{Bird+2016,Sasaki+2018}.

Our focus is on the formation of merging binary BHs through both
CE evolution and stable MT. The majority of predicted binary BHs
formed through CE evolution or stable MT in an isolated binary
involve an initial phase of stable MT. This removes the hydrogen envelope
of the more massive star and leads to the formation of a wide single degenerate
binary \citep{Dominik+2012, Langer+2020}. When the secondary evolves to fill its
Roche lobe, depending on its response to mass loss and the mass ratio of the
system it can undergo a CE phase which ejects the envelope of the
secondary at the cost of hardening the binary \citep[cf.][]{Belczynski+2016}.
If the second phase of MT is stable, \citet{vandenHeuvel+2017} argued
that depending on the mass ratio of the system a phase of non-conservative mass
transfer can harden the binary without the need for CE evolution, potentially
allowing for the formation of compact object binaries that can merge within a
Hubble time. Recent studies have claimed that the contribution from stable mass
transfer to the observed sample of merging binary BHs can be comparable or even
larger than those formed through CE evolution \citep{Neijssel+2019,
Bavera+2020b}.

However, binary evolution models have large uncertainties associated
with the CE
phase (see \citealt{Ivanova+2013} for a recent review). Important open
issues include the unknown efficiency of CE evolution
\citep{Zorotovic+2010,Davis+2012,ToonenNelemans2013}, the mass boundary at
which CE evolution terminates 
\citep[the core--envelope boundary;][]{Han+1994,DewiTauris2000,Ivanova2011}, and the
actual conditions for the onset of CE evolution \citep{HjellmingWebbink1987,
Soberman+1997, Pavlovskii+2017}. Observational and theoretical constraints for
the outcome of CE evolution in massive stars are also limited, as the vast
majority of observed post-CE systems as well as multi-dimensional hydrodynamical simulations of CE
evolution correspond to low and intermediate-mass stars (see
\citealt{IaconiDeMarco2019} for a recent compilation of CE simulations and
observed post-CE systems). 

One uncertainty of CE evolution that is studied in population synthesis
calculations is the impact of the evolutionary stage of the star at the onset of
CE evolution. \citet{Belczynski+2010} proposed that in low metallicity
environments a larger number of massive stars are expected to survive CE
evolution leading to an enhanced formation rate of merging binary BHs.
This is argued as a consequence of halted expansion after
core-hydrogen burning, with stellar models undergoing core-helium burning as
blue supergiants and covering a larger range of radii after core-helium
ignition. Interaction after core-helium ignition, argues
\citet{Belczynski+2010}, favors the ejection of the envelope as at that point a
well defined core--envelope structure is formed. In various population synthesis
codes this argument is encoded in Pessimistic and Optimistic options for CE
survival. In the Pessimistic approach all systems undergoing CE
evolution before core-helium ignition are assumed to merge during CE evolution,
while the Optimistic approach allows for their survival. Recently,
\citet{Klencki+2020a} and \citet{Klencki+2020b} have argued that this approach
likely overestimates the number of merging binary BHs formed, as most of the
systems predicted to survive CE in the Optimistic scenario would initiate CE
evolution with a radiative envelope that has a binding energy too large to eject
through an inspiral.

The aim of this work is to study in detail how the evolutionary stage of the
stellar progenitor of a BH affects MT stability and the outcomes of
CE evolution. We do this by computing detailed models with the \texttt{MESA}
code \citep{Paxton+2011,Paxton+2013,Paxton+2015,Paxton+2018,Paxton+2019} of a
low metallicity massive star with a BH companion, meant to reproduce the
evolution of a binary system after the formation of the first BH. {Recent work
on the formation of binary BHs using detailed binary evolution
models (in contrast to rapid population synthesis calculations) has been
presented by \citet{EldridgeStanway2016} and \citet{Pavlovskii+2017}, covering a broad range of
donor masses and metallicities. Our work, in contrast, focuses on a single donor
mass of $30M_\odot$ at a metallicity of $Z_\odot/10$, but with a
large resolution in initial mass ratio and period. This allows us to study in
detail how the stability of MT and the outcomes of CE evolution are
affected by the evolutionary stage of the donor}. For these
calculations, we have made improvements to the commonly used MT
prescription of \citet[][hereafter \citetalias{KolbRitter1990}]{KolbRitter1990} and implemented a method to model CE
evolution which self-consistently determines the core--envelope boundary. Our
methods are described in Sec.~\ref{sec:methods} and we present our results in
Sec.~\ref{sec:results}. We conclude by discussing our results in Sec.~\ref{sec:discussion}. All input files necessary to reproduce our simulations as
well as associated data products are available for download at
\href{https://doi.org/10.5281/zenodo.4106318}{doi.org/10.5281/zenodo.4106318}.

\section{Methods} \label{sec:methods}
We perform our simulations using version {15140} of the \texttt{MESA} stellar
evolution code. Our models consist of binary systems with a $30M_\odot$ donor star and a BH
companion. The objective of
these simulations is to model a MT event happening after the
formation of the first BH in the system, which is argued to be the most
important MT phase where a CE can lead to the formation of a merging
binary BH \citep[cf.][]{Dominik+2012}. Since we do not model the previous
evolution of the system, we approximate the starting state of the donor star in the system as a
zero-age main-sequence (ZAMS) star of its mass. Evolution is computed until
either core carbon is depleted or a merger between the donor star and the BH
happens.

MT rates are computed using the method described in
Sec.~\ref{section:MT}, while the actual accretion rate into the BH is limited
to its Eddington rate as described in \citet{Marchant+2017}. Our model for CE
evolution is described in Sec.~\ref{sec:CE}. Mass loss either
due to stellar winds from the donor, or from mass ejected from the vicinity of
the BH by MT above the Eddington limit is assumed to take an amount
of angular momentum corresponding to the specific orbital angular momentum of each
component. We also account for the possibility of a star growing to the point
that not only it overfills its Roche lobe, but also overflows its outer
Lagrangian point. Whenever this happens we account for the loss of mass and angular
momentum as described in Sec.~\ref{sec:mout}. Angular momentum loss from GW radiation
is also included \citep{Peters1964}, but does not play a role in the $\sim\mathrm{Myr}$
timescales of evolution of our models. Our simulations do not take into account
stellar rotation or spin--orbit coupling.

Our stellar models are computed at a low metallicity of $Z=Z_\odot/10$ where
we take the relative metal mass fractions {from \citet{GrevesseSauval1998} and
take the solar abundance to be} $Z_\odot=0.0142$ \citep{Asplund+2009}. Opacities are computed using opacity tables from the
OPAL project \citep{IglesiasRogers1996}, together with the low temperature
opacity tables of \citet{Ferguson+2005}. The equation of state used by
\texttt{MESA} consists of a combination of {OPAL
\citep{RogersNayfonov2002}},
HELM \citep{TimmesSwesty2000}, PC \citep{PotekhinChabrier2010} and SCVH
\citep{Saumon+1995}.
Nuclear reaction rates are taken {in order of preference from
\citet{Cyburt+2010} and \citet{Angulo+1999}}.

Convection is modeled using the mixing length
theory of \citet{Bohm-Vitense1958} as described by \citet{CoxGiuli1968}, using a
mixing length parameter $\alpha_\mathrm{MLT}=2$. Convective regions are determined using the Ledoux
criterion \citep{Ledoux1947}. We include overshooting of the hydrogen burning convective core
using a step-overshooting scheme, extending the size of the convective core by
$\alpha_\mathrm{ov}=0.335$ pressure scale heights following the calibration of \citet{Brott+2011}.
Overshooting from other convective regions is not well understood, {so for
convective cores after the main-sequence we 
include a small amount of exponential overshooting \citep{Herwig2000}, given by a length scale for
exponential decay of the mixing coefficient of $f=0.01$.\footnote{In practice,
since formally at the edge of a convective region there is no mixing,
overshooting in \texttt{MESA} is specified by a pair of variables $f$ and $f_0$,
where $f_0$ determines a distance in units of pressure scale heights into the
convective region, from which the mixing coefficient from overshooting is
computed. For step overshooting we take $f=0.345$ and $f_0=0.01$, while for
exponential overshooting we use $f=0.01$ and $f_0=0.005$}} We also include
semiconvective mixing \citep{Langer+1983} and thermohaline mixing
\citep{Kippenhahn+1970}. {Following \citet{Schootemeijer+2019}, we
adopt a large value for the efficiency parameter for semiconvection $\alpha_\mathrm{sc}=100$, while for thermohaline mixing we adopt an efficiency parameter of
unity}.

Stellar winds are accounted for using a combination of different mass loss rate
prescriptions as described by \citet{Brott+2011}. This includes the line-driven
mass loss rates derived by \citet{Vink+2001} for stars with a hydrogen surface
mass fraction $X>0.7$ and the
Wolf--Rayet mass loss rate of \citet{Hamann+1995}, scaled by a factor of $10$ to
account for wind clumping \citep{Yoon+2010}, for $X<0.4$.
For surface hydrogen mass fractions
between $X=0.7$ and $X=0.4$ we interpolate between the rate of
\citet{Vink+2001} and \citet{Hamann+1995} to provide a continuous transition.
At temperatures below that of the bi-stability jump (as derived by
\citealt{Vink+2001}) we take the maximum between the rate of
\citet{NieuwenhuijzendeJager1990} and the one resulting from the combination of
the \citet{Vink+2001} and \citet{Hamann+1995} rates we just described.
Additionally, we scale the \citet{NieuwenhuijzendeJager1990} {rate} by the same factor
$(Z/Z_\odot)^{0.85}$ predicted by \citet{Vink+2001} for line-driven winds. This significantly lowers the mass loss rates of red-supergiants in low
metallicity environments, although observations suggest there is only a weak
dependence on metallicity \citep{vanLoon+2005,Goldman+2017}. Nevertheless we
keep this scaling for the sake of comparison, as it is commonly used in
population synthesis calculations.

\subsection{Mass transfer} \label{section:MT}

We model MT using an extension of the method developed by
\citetalias{KolbRitter1990}, which accounts for MT from an extended
atmosphere as described in \citet[][hereafter \citetalias{Ritter1988}]{Ritter1988} as well as the case where
optically thick regions overflow the Roche lobe of the donor. In particular,
this does not assume that the photosphere of the star operates as a hard rim,
potentially leading to significant amounts of overflow during MT.
Here we describe in detail the method of \citetalias{KolbRitter1990} and our
modifications to it, which include the possibility of outflows from the outer
Lagrangian point of the donor. {For the case of overflow from
optically thick regions, we also discuss similarities of our method with that of
\citet{PavlovskiiIvanova2015}.}

\begin{figure}
   \includegraphics[width=\columnwidth]{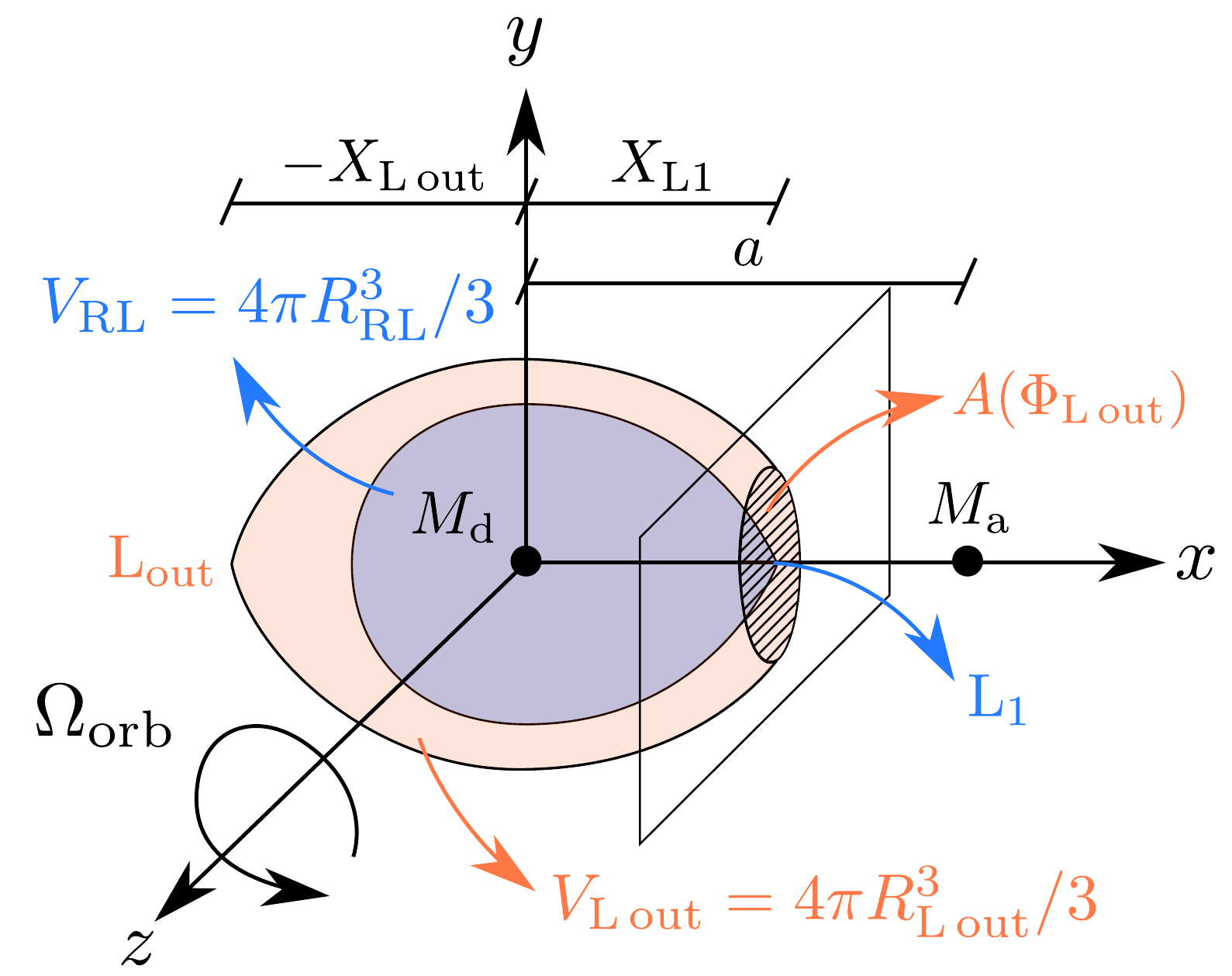}
   \caption{Definition of the coordinate system and variables used in the
   calculation of MT. The coordinate system is centered in the donor
   star, with the $x$-axis defined as the line joining the donor and the accretor,
   with the accretor located at $x=a$ where $a$ is the orbital separation. The
   $z$-axis is oriented along the direction of the orbital angular momentum of the
   system, while the $y$-axis completes a standard right-handed oriented Cartesian
   set of coordinates. The Lagrangian point located behind the donor is either
   L$_3$ or L$_2$, depending on whether the donor is more or less massive than the
   accretor respectively.}\label{fig:coords}
\end{figure}

\subsubsection{The Roche potential}

The computation of the MT rate depends on the Roche potential $\Phi$,
which is taken to be that of two point masses in a circular orbit:
\begin{eqnarray}
   \Phi = \frac{G M_\mathrm{d}}{a}\left\{-\frac{1}{\hat{r}_\mathrm{d}}-\frac{q}{\hat{r}_\mathrm{a}}-\frac{q+1}{2}\left[(\hat{x}-\hat{x}_\mathrm{cm})^2 +
   \hat{y}^2\right]\right\},
\end{eqnarray}
where $M_\mathrm{d}$ is the mass of the donor and the mass ratio is defined as
$q\equiv M_\mathrm{a}/M_\mathrm{d}$ with $M_\mathrm{a}$ being the mass of the accretor.
We use a coordinate system centered on the donor star (see Fig.~\ref{fig:coords}), denoting distances
normalized by the orbital separation $a$ as $\hat{x}=x/a$; $\hat{r}_\mathrm{d}$ and
$\hat{r_\mathrm{a}}$ are the normalized distances of a point to the center of each
star, and $\hat{x}_\mathrm{cm}=q/(q+1)$ indicates the $x$-coordinate of the center of
mass of the system. The $x$-coordinate of the first Lagrangian point is given by
$X_\mathrm{L1}$. The $x$-coordinate of the outer Lagrangian point of the donor
(which can be either L$_2$ or L$_3$ depending on the mass ratio) is given by $X_\mathrm{L\,out}$, which is defined negative.
The Roche lobe radius of the system $R_\mathrm{RL}$ is defined as the volume
equivalent radius of the region below the equipotential at the first Lagrangian
point $\Phi_\mathrm{L1}$, and we use the fitting formula from \citet{Eggleton1983}
to approximate it,
\begin{eqnarray}
   \frac{R_\mathrm{RL}}{a}=\frac{0.49 q^{-2/3}}{0.6q^{-2/3} +
   \ln\left(q^{-1/3}+1\right)}.\label{equ:egg}
\end{eqnarray}

\begin{figure}
   \includegraphics[width=\columnwidth]{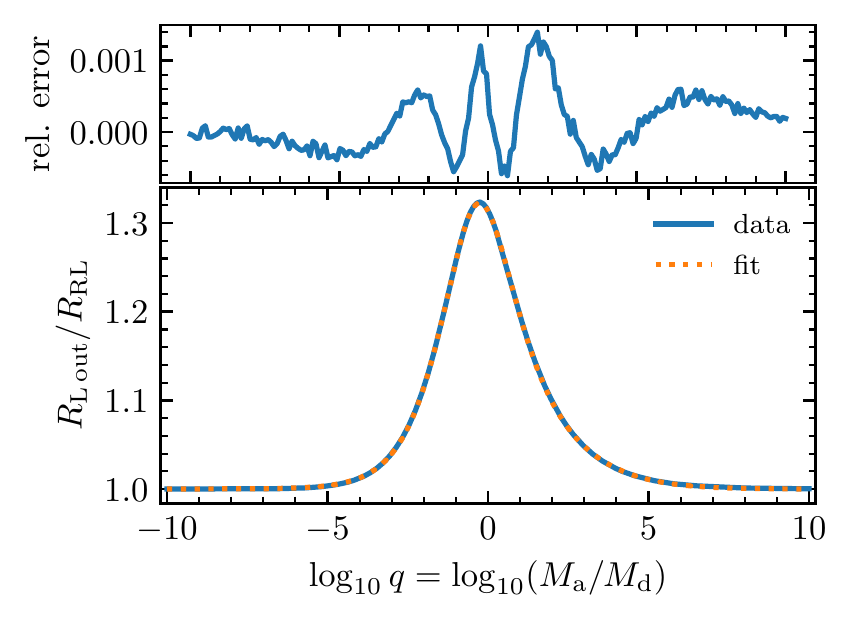}
   \caption{Ratio between the volume equivalent radius associated to the outer
   Lagrangian point of the donor and its Roche lobe radius. Bottom panel shows
   the result from numerical integration of the Roche potential together with
   the fit given in Eq.~\eqref{equ:rout}. Top panel shows the relative error
   between the fit and the data, with the small variations being due to the
   finite precision of the volume calculation.}\label{fig:rout_fit}
\end{figure}

Since we consider stars with significant overflow,
potentially reaching the outer Lagrangian point of the donor, we also define
volume-equivalent radii beyond the equipotential of L$_1$ by dividing
space with a plane crossing L$_1$ which is perpendicular to the line joining
both stars, as depicted in Fig.~\ref{fig:coords}. We refer to this plane as the
L$_1$ plane. The volume equivalent radius
for an arbitrary value of $\Phi$ is then defined as $V(\Phi)=4\pi R(\Phi)^{3}/3$,
where $V(\Phi)$ is the volume contained within the equipotential $\Phi$ and on
the side of the L$_1$ plane where the donor is located. We care in particular
about the radius associated to the equipotential of the outer Lagrangian point of
the donor, for which we find the following equation provides a fit with an error
$<0.15 \%$ in the range $-10 < \log_{10} q < 10$:\footnote{To construct this fit we numerically computed both $R_\mathrm{RL}$
and $R_\mathrm{L\,out}$ for a wide range of mass ratios, without using the
\citet{Eggleton1983} approximation as it has an error $\sim 1\%$ for some mass
ratios. However, for consistency, when we evaluate $R_\mathrm{L\,out}$ in the
simulations in this work we use Eq.~\eqref{equ:rout} together with the fit of
\citet{Eggleton1983} to ensure that $R_\mathrm{L\,out}>R_\mathrm{RL}$.}
\begin{eqnarray}
   \begin{split}
   \label{equ:rout}
   \frac{R_\mathrm{L\,out}}{R_{\rm
      RL}}&=&1+\frac{2.74}{1+[(\ln q +
   1.02)/\sigma]^2}\times\frac{1}{7.13+q^{-0.386}},\qquad \\
      \sigma &=&
      \frac{49.4}{12.2+q^{0.208}}.\qquad\qquad\qquad\qquad\qquad
   \end{split}
\end{eqnarray}
Figure~\ref{fig:rout_fit} shows the value of $R_\mathrm{L\,out}/R_\mathrm{RL}$ for
different values of the mass ratio. In the limit of $\log_{10}q\rightarrow\pm
\infty$ we have that $R_\mathrm{L\,out}/R_\mathrm{RL}\rightarrow 1$, while in the
range $-1 < \log_{10}q < 1$, which is typical of interacting binary stars, we
find that $R_\mathrm{L\,out}/R_\mathrm{RL}$ is between $\sim 1.2$--$1.3$. Thus, a donor
star growing beyond $\sim 20\%$--$30\%$ percent of its Roche lobe radius is
expected to also fill its outer Lagrangian point.

\subsubsection{Mass transfer through the L$_1$ point}

Restricting the discussion first to flows through the L$_1$ point, the MT
rate can be computed as an integral over the L$_1$ plane,
\begin{eqnarray}
   \dot{M}_\mathrm{mt,{\rm L}1} = \int \rho v \,\mathrm{d}A, \label{equ:intmt}
\end{eqnarray}
where $\rho$ and $v$ are the density and velocity of the fluid in the L$_1$ plane. The
flow is assumed to be steady in which case the Bernoulli equation is satisfied,
\begin{eqnarray}
   \frac{1}{2}v_\mathrm{f}^2+\int_\mathrm{i}^\mathrm{f} \frac{\mathrm{d} P}{\rho} + \Phi_\mathrm{f} =
   \frac{1}{2}v_\mathrm{i}^2 + \Phi_\mathrm{i},\label{equ:ber}
\end{eqnarray}
where the integral is done along a streamline of the fluid with
{$\mathrm{i}$ and $\mathrm{f}$ denoting an initial and final point
along a streamline}.
Following the work of \citet{LubowShu1975}, the velocity near the
Lagrangian point is expected to be equal to the sound speed. Depending on
whether the photosphere of the star is inside the Roche lobe or has expanded
beyond it, the flow is respectively assumed to be isothermal \citepalias{Ritter1988}
or adiabatic \citepalias{KolbRitter1990}. Combining either the adiabatic or
isothermal approximation with Eq.~\eqref{equ:ber} allows the computation of
$\rho$ and $v$ in the vicinity of the Lagrangian point as a function of the Roche
potential $\Phi$.

Either in the isothermal or adiabatic approximation, 
the surface integral in Eq.~\eqref{equ:intmt} can be expressed as
an integral over the potential,
\begin{eqnarray}
   \dot{M}_\mathrm{mt,{\rm L}1} = \int \rho v \frac{\mathrm{d} A}{\mathrm{d} \Phi} \mathrm{d}\Phi
   \label{equ:intmt},
\end{eqnarray}
where $A(\Phi)$ is the area in the L$_1$ plane below the equipotential $\Phi$.
Figure~\ref{fig:coords} illustrates the area $A(\Phi_\mathrm{L\;out})$. The $\mathrm{d}A/\mathrm{d}\Phi$ term describes how the area enclosed by an equipotential in the
L$_1$ plane changes with $\Phi$, and encodes the properties of the Roche
geometry. Along the L$_1$ plane, given the symmetry of $\Phi$ and the L$_1$ point
being a local minimum, the Roche potential can be approximated as
\begin{eqnarray}
      \Delta \Phi \simeq  C_1 y^2 + C_2 z^2 +C_3 y^4 + C_4 y^2 z^2 + C_5
      z^4,\label{equ:taylor}
\end{eqnarray}
where $\Delta \Phi\equiv\Phi-\Phi_{\mathrm{L}1}$. As shown in Appendix~\ref{app:dadphi}, $C_3=C_5=C_4/2$ and the value of $\mathrm{d} A/\mathrm{d}\Phi$ can be computed up to first order in
$\Delta\Phi$ as
\begin{eqnarray}
   \begin{split}
   \frac{\mathrm{d} A}{\mathrm{d} \Phi} =
   \frac{\pi}{\sqrt{C_1C_2}}\left(1 -
      \frac{C_3(3C_1^2+2C_1C_2+3C_2^2)}{4C_1^2C_2^2}\Delta \Phi\qquad\right.\\
      \left.+\mathcal{O}\left[\left(\Delta \Phi\right)^2\right]\right)
   \end{split}\label{equ:dadphi_f}
\end{eqnarray}
Both \citetalias{Ritter1988} and \citetalias{KolbRitter1990} approximate $\mathrm{d}A/\mathrm{d}\Phi$ as a
constant, but as we want to consider the case where the star can significantly
overfill its Roche lobe, we include the term that goes as $\Delta \Phi$ when
computing MT in the optically thick case. \citetalias{Ritter1988} and
\citetalias{KolbRitter1990} also rely on fits to the coefficients $C_1$ and $C_2$ as
a function of mass ratio. To avoid relying on fitting formulae with a
limited range of validity we directly compute the location of the L$_1$ point, $X_\mathrm{L1}$, and compute all $C_i$ coefficients explicitly from the Roche
potential.\footnote{\citetalias{Ritter1988} define a function $F(q)$ in their Eq.~(A8) which contains the product $C_1C_2$, and provide a fit for this function. The expression provided by \citetalias{Ritter1988}
has a typo which is corrected in Eq.~(A3) of \citetalias{KolbRitter1990}. Despite this typographical error in
\citetalias{Ritter1988}, we have verified that the fit to $F(q)$ provided in Eq.~(A9) of \citetalias{Ritter1988} matches the
correct expression.}
{The variation of $\mathrm{d}A/\mathrm{d}\Phi$ with increasing overflow has also been taken into
account for optically thick Roche lobe
overflow (RLOF) by \citet{PavlovskiiIvanova2015}, but using
tables with numerical integrations of the area in the L$_\mathrm{1}$ plane instead of a higher-order analytical approximation.}

\subsubsection{Optically thin mass transfer ($R_\mathrm{ph}<R_{\rm
RL}$)}\label{sec:thin}
For the case of MT from a star with a photosphere radius $R_\mathrm{ph}$ below its
Roche lobe radius $R_\mathrm{RL}$, \citetalias{Ritter1988}
takes the flow to have a constant temperature equal to $T_\mathrm{eff}$,
in which case the relevant sound speed is the isothermal one,
\begin{eqnarray}
   v_\mathrm{th}^2=\left(\frac{\partial P}{\partial \rho}\right)_T = \frac{kT_{\rm
   eff}}{\mu m_h},\label{equ:visot}
\end{eqnarray}
where $k$ is the Boltzmann constant, $\mu$ is the mean molecular weight and $m_h$ is one atomic
mass unit. This sound speed corresponds to the equation of an
ideal gas plus radiation, 
\begin{eqnarray}
   P=\frac{\rho k T_\mathrm{eff}}{\mu m_h} + \frac{a_\mathrm{rad}T_\mathrm{eff}^4}{3},
   \label{equ:igpr}
\end{eqnarray}
where $a_\mathrm{rad}$ is the radiation constant. {We consider
streamlines that start with negligible velocity from the photosphere of the star
by taking $v_\mathrm{i}=0$ and $\Phi_\mathrm{i}=\Phi_\mathrm{ph}$ in Eq.~\eqref{equ:ber},
where $\Phi_\mathrm{ph}$ is the value of the Roche potential at the photosphere}.
Combining Eq.~\eqref{equ:ber}, Eq.~\eqref{equ:visot} and Eq.~\eqref{equ:igpr} one can solve
for the density of the flow at the L$_1$ point,
\begin{eqnarray}
   \rho_{{\rm L}1} = \frac{\rho_\mathrm{ph}}{\sqrt{e}}\exp\left(\frac{\Phi_{\rm
   ph}-\Phi_{{\rm L}1}}{v_\mathrm{th}^2}\right),\label{equ:rhothin}
\end{eqnarray}
where the $\mathrm{ph}$ and ${\rm L}1$ subscripts indicate properties at the
photosphere and the Lagrangian point.

Eq.~\eqref{equ:rhothin} can be evaluated in the vicinity of the L$_1$ point to
express the variation in density across the L$_1$ plane as a function of the
variation in $\Phi$ and $v$,
\begin{eqnarray}
   \mathrm{d}\rho = -\frac{\rho_{{\rm L}1}}{v_{\rm
   th}^2}\left(\mathrm{d}\Phi+\frac{(\Phi_\mathrm{ph}-\Phi_{{\rm L}1})\mathrm{d} v}{v_{\rm
   th}}\right).\label{equ:drhothin}
\end{eqnarray}
As for a steady flow the L$_1$ point
operates as a nozzle, acceleration perpendicular to the flow is expected to
vanish on the L$_1$ plane,
\begin{eqnarray}
   \mathrm{d}P=-\rho\mathrm{d}\Phi \rightarrow \mathrm{d}\rho=\frac{\rho}{v_\mathrm{th}^2}\mathrm{d}\Phi, \label{equ:drho}
\end{eqnarray}
which combined with Eq.~\eqref{equ:drhothin} results in $\mathrm{d}v=0$ (and thus
$v=v_\mathrm{th}$) in the vicinity of L$_1$.

These results can be used to evaluate the MT rate using Eq.~\eqref{equ:intmt}. Transforming the integral into one over density with Eq.~\eqref{equ:drho} results in
\begin{eqnarray}
   \dot{M}_\mathrm{thin}=-\int_{\rho_{{\rm L}1}}^0 v_\mathrm{th}^3\frac{\mathrm{d}A}{\mathrm{d}\Phi}
   \mathrm{d}\rho.
\end{eqnarray}
As the density decays exponentially with the potential, see
Eq.~\eqref{equ:rhothin}, only the regions in the very vicinity of L$_1$ are expected
to contribute to the MT rate, so we approximate $\mathrm{d}A/\mathrm{d}\Phi$ by its
value at L$_1$ to obtain
\begin{eqnarray}
   \dot{M}_\mathrm{thin}=\dot{M}_{0,\rm thin} \exp\left(\frac{\Phi_\mathrm{ph}-\Phi_{{\rm L}1}}{v_\mathrm{th}^2}\right),
\end{eqnarray}
where $M_{0,{\rm thin}}$ is the MT rate for a star just filling its
Roche lobe,
\begin{eqnarray}
   \dot{M}_{0,\rm thin} =\frac{\rho_\mathrm{ph}v_{\rm
   th}^3}{\sqrt{e}}\left(\frac{\mathrm{d}A}{\mathrm{d}\Phi}\right)_{{\rm L}1}.
\end{eqnarray}

The factor that remains to compute the MT rate in an evolutionary
code is the difference $\Phi_{{\rm ph}}-\Phi_{{\rm L}1}$. \citetalias{Ritter1988} opt
to use the derivative of the potential with respect to volume equivalent radii
$r_V$ covered by different equipotentials, and approximate
\begin{eqnarray}
   \Phi_\mathrm{ph}-\Phi_{{\rm L}1} = \left(\frac{\mathrm{d}\Phi}{\mathrm{d} r_V}\right)_{{\rm L}1}\left(R_\mathrm{
      RL}-R_\mathrm{ph}\right).
\end{eqnarray}
We follow instead the method of \citet{Jackson+2017} who developed an expansion
of the Roche potential around the donor as a function of the equivalent radius $r_V$,
\begin{eqnarray}
   \begin{split}
   \Phi(r_V)=-\left(\frac{GM_a}{a}+\frac{GM_a}{2a(M_d+M_a)}\right)\qquad\qquad\qquad\qquad\qquad\\
      -\frac{GM_d}{r_V}\left[1+a_1\left(\frac{r_V}{a}\right)^3+a_2\left(\frac{r_V}{a}\right)^6+\mathcal{O}\left(\frac{r_V}{a}\right)^9\right],
   \end{split}
\end{eqnarray}
with $a_1$ and $a_2$ given by
\begin{eqnarray}
   \begin{split}
      a_1&=&\frac{1}{3}\left(\frac{M_d+M_a}{M_d}\right),\qquad\qquad\qquad\qquad\qquad\quad\;\;\\
      a_2&=&\frac{4}{45}\left(\frac{(M_d+M_a)^2+9M_a^2+3M_a(M_d+M_a)}{M_d^2}\right).
   \end{split}
\end{eqnarray}
Using this we compute $\Phi_{{\rm L}1}$ with the Roche lobe radii
from Eq.~\eqref{equ:egg}, and compute $\Phi_\mathrm{ph}$ using the photospheric
radius of our stellar model.

In all our simulations only a small fraction of the total
MT comes from the contribution of this thin MT rate, but it
still operates as physically motivated mechanism to smoothly turn on MT which helps prevent
numerical instabilities in the calculations when a star initiates RLOF.

\subsubsection{Optically thick mass transfer ($R_\mathrm{ph}>R_\mathrm{RL}$)}
\label{sec:thick}

Following \citetalias{KolbRitter1990}, if layers of the star below the photosphere
are overflowing the Roche lobe, then the MT rate is computed as
\begin{eqnarray}
   \dot{M}_\mathrm{mt,{\rm L}1} = \dot{M}_{0,\rm thin}+\dot{M}_{\rm thick},
\end{eqnarray}
where $\dot{M}_{\rm thick}$ is the contribution to the integral in Eq.~\eqref{equ:intmt} from the overflowing optically thick regions,
\begin{eqnarray}
   \dot{M}_\mathrm{thick}=\int_{\Phi_{{\rm L}1}}^{\Phi_\mathrm{ph}} \rho v \frac{\mathrm{d}
   A}{\mathrm{d} \Phi} \mathrm{d}\Phi.\label{equthick}
\end{eqnarray}
To calculate the density and velocity of the flow at L$_1$ we again make use of the
Bernoulli equation, Eq.~\eqref{equ:ber}, by assuming the flow moves parallel to the
equipotential surfaces and is adiabatic such that the pressure and sound speed are
given {along a streamline} by
\begin{eqnarray}
   P=k\rho^{\Gamma_1},\quad v_s^2=\frac{\Gamma_1P}{\rho}. \label{equstate}
\end{eqnarray}
The constant $k$ can be computed in terms of a reference pressure and density
{for each streamline},
$k=P_0/\rho_0^{\Gamma_1}$. {This assumes $\Gamma_1$ is constant along
the streamline; \citet{PavlovskiiIvanova2015} have considered the case of a
variable $\Gamma_1$ and found the resulting MT rates are only
modified by a few percent.}

We assume that along each equipotential surface, far
from the Lagrangian point, the fluid is near hydrostatic equilibrium ($\mathrm{d}P_0\simeq-\rho \mathrm{d}\Phi$,
and $v\ll v_s$) and take the value of $P_0$ and $\rho_0$ there to
compute $k$. Combining Eq.~\eqref{equstate} with Eq.~\eqref{equ:ber},
and taking
$\Phi_\mathrm{i}=\Phi_\mathrm{f}$, $v_\mathrm{i}=0$ and $v_\mathrm{f}=v_{s}$, one can compute
the density of the flow in the L$_1$ plane for a given value of $\rho_0$,
\begin{eqnarray}
   \rho=\rho_0\left(\frac{2}{\Gamma_1+1}\right)^{1/(\Gamma_1-1)}.\label{equ:rhoL}
\end{eqnarray}
Combining this density with Eq.~\eqref{equthick}
gives the contribution to the MT rate from the overflowing optically
thick layers as
\begin{eqnarray}
   \dot{M}_{\rm
   thick}=\int_{P(R_\mathrm{RL})}^{P(R_{\rm
   ph})}\left(\frac{P_0}{\rho_0}\right)^{1/2}F_3(\Gamma_1)\frac{\mathrm{d}A}{\mathrm{d}\Phi}\mathrm{d}P_0,
   \label{equ:thick2}
\end{eqnarray}
where
\begin{eqnarray}
   F_3(\Gamma_1)=\Gamma_1^{1/2}\left(\frac{2}{\Gamma_1+1}\right)^{{(\Gamma_1+1)}/[{2(\Gamma_1-1)}]}.
\end{eqnarray}
In our calculations we take $P_0$ and $\rho_0$ to be the density in the
overflowing layers of our hydrostatic model, from which we also obtain
$\Gamma_1$. In this case $P(R_\mathrm{RL})$ and $P(R_\mathrm{ph})$ are the pressure of
the stellar model in the layer were $R=R_\mathrm{RL}$ and $R=R_\mathrm{ph}$
respectively.

Equation~\eqref{equ:thick2} is exactly as derived by \citetalias{KolbRitter1990}, but
in our calculations there are two important differences on how we compute it.
First, \citetalias{KolbRitter1990} make the assumption of an ideal gas, replacing the
$P_0/\rho_0$ ratio in Eq.~\eqref{equ:thick2} by $kT/\mu m_h$. In layers
dominated by radiation pressure this underestimates the mass loss rate
significantly, so we do not make this assumption and compute $P_0/\rho_0$ from
our stellar model. The second change is that we do not assume $\mathrm{d}A/\mathrm{d}\Phi$ to be
constant, but instead evaluate it using Eq.~\eqref{equ:dadphi_f}, with $\Delta
\Phi$ being computed in the overflowing layers of the model as
\begin{eqnarray}
   \Delta \Phi = \int_{P(R_\mathrm{RL})}^{P}\frac{\mathrm{d}P_0}{\rho_0}.
\end{eqnarray}

\subsubsection{Mass loss from the outer Lagrangian point of the
donor}\label{sec:mout}

Models of interactive massive stars using the prescription of
\citetalias{KolbRitter1990} can exhibit radii much larger than their Roche lobe
radii.
In particular in the simulations presented in
this work involving a $30M_\odot$ donor we find cases where the outer
Lagrangian point would be overflowed ($R>R_\mathrm{L\;out}$).
For these cases we need a model for the overflow from this outer Lagrangian point,
which is L$_3$ if
$M_\mathrm{d}>M_\mathrm{a}$ and L$_2$ otherwise.

The entire method described in Sec.~\ref{sec:thin} and Sec.~\ref{sec:thick} can
be applied for an outflow through the outer Lagrangian point, by
replacing the Roche lobe radius with $R_\mathrm{L\;out}$ using Eq.~\eqref{equ:rout}
and evaluating the $\mathrm{d}A/\mathrm{d}\Phi$ factor at the outer Lagrangian point instead of
L$_1$. However, an assumption needs to be made as to what happens to this outflow.
We assume the material is ejected from the system carrying the
specific angular momentum that corresponds to the outer Lagrangian point,
\begin{eqnarray}
   j_{\mathrm{L\;out}} = \frac{2\pi}{P_\mathrm{orb}}a^2\left(\frac{M_a}{M_a+M_d}-\hat{X}_\mathrm{L\;out}\right)^2.\label{equ:jout}
\end{eqnarray}
In order for this to be satisfied the outflow needs to have a velocity much
larger than the orbital velocity, otherwise tidal forces from the binary can
modify the angular momentum content of ejected material.
This has been studied in the
case of overcontact binary systems that overflow the L$_2$ point
\citep{Shu+1979,Pejcha2014,Pejcha+2016}, or have their stellar winds torqued as they
accelerate \citep{macLeodLoeb2020}.
If the energy of the
outflowing material is not sufficient to unbind it from the binary, it can also
accumulate in a circumbinary disk.

\begin{figure}
   \includegraphics[width=\columnwidth]{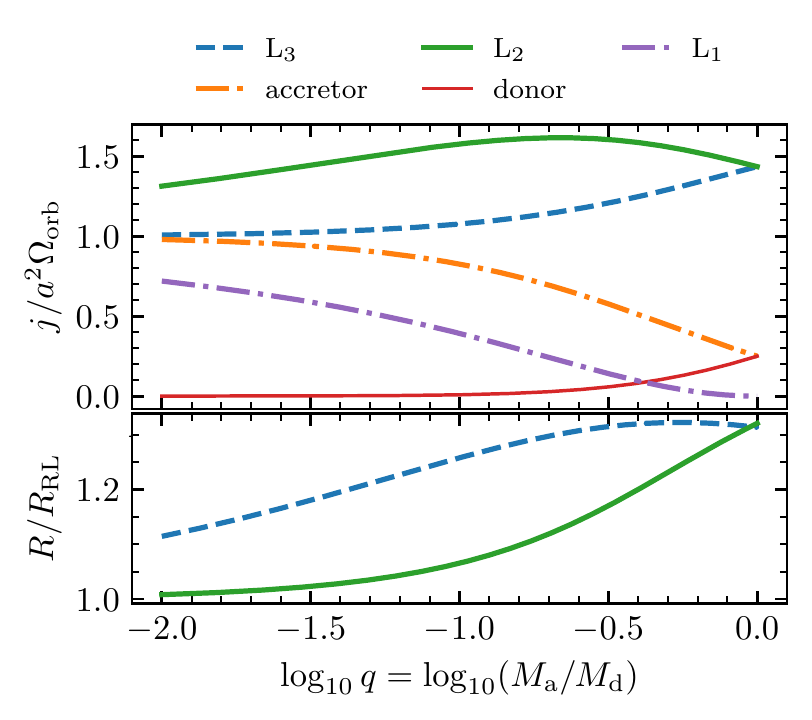}
   \caption{\textit{Top:} Specific angular momentum in units of $a^2\Omega_\mathrm{orb}$ for the
   donor, the accretor, and the first three Lagrangian points as a function of
   mass ratio. {\textit{Bottom:} Volume equivalent radii for the donor corresponding
   to the L$_2$ and L$_3$ equipotentials. Volume equivalent radius for L$_2$ is
   taken from the fit of \citet{Marchant+2016}, which due to small errors in the
   fit results in a slightly larger value than the radius for the L$_3$
   equipotential at $q=1$.
   }}\label{fig:specificj}
\end{figure}

Owing to {these uncertainties}, we also
experiment with models where the specific angular momentum in Eq.~\eqref{equ:jout} is increased by factors of $4$ and $9$, corresponding to
points co-rotating with the orbit at $2$ and $3$ times the distance of
L$_\mathrm{out}$ from the center of mass. {However, any of our simulations that
undergo stable MT while overflowing L$_2$ or L$_{3}$ should be considered with care,
since this is an extreme regime where three dimensional hydrodynamic effects
become important.}

\subsubsection{Outflows from L$_2$ when
$M_\mathrm{a}/M_\mathrm{d}<1$}\label{sec:L2out}
An alternative approach to the problem of large overflow in simulations was taken by
\citet{Misra+2020}. Considering systems with neutron star
accretors, they computed the equivalent radii associated to the
full L$_{2}$ equipotential (including both the donor and the accretor),
and assumed that whenever a stellar model exceeded
this radius then MT becomes unstable owing to the formation of an outflow with a
high specific angular momentum. 
However, since for a compact object accretor
there is not a hydrostatic structure within the Roche lobe of the accretor, the
volume associated to the full L$_{2}$ equipotential cannot be compared
directly to the radii of a hydrostatic stellar model in order to assess overflow of the outer
Lagrangian points.

Nevertheless, in cases where $M_\mathrm{d}>M_\mathrm{a}$ some of
the material streaming through L$_{1}$ could potentially be ejected from the
vicinity of L$_{2}$ before the donor overflows $L_{3}$. In this case our
simulations would underestimate angular momentum loss from the system, since the
specific angular momentum associated to L$_{2}$ can be up to $47\%$ larger
than that of L$_{3}$, as is shown in Fig.~\ref{fig:specificj}. Fig.~\ref{fig:specificj}
also indicates the volume equivalent radii for the donor associated to the L$_3$
and L$_2$ equipotentials when $M_\mathrm{a}/M_\mathrm{d}<1$, where for 
the donor's L$_2$ equivalent radius we use the fit of \citet{Marchant+2016},
\begin{eqnarray}
   \frac{R_{\mathrm{L}_2}}{R_\mathrm{RL}}=1+0.299 \tan^{-1}(1.84
   q^{0.397})q^{0.520}.
\end{eqnarray}
Although for a broad range of mass ratios the donor can expand significantly
before filling out the L$_3$ equipotential, much less expansion beyond the Roche
lobe is required to fill out L$_2$.

In order to assess potential uncertainties during phases of large overflow,
where material streaming through L$_1$ could be ejected from the vicinity of
L$_2$ when $M_\mathrm{a}/M_\mathrm{d}<1$, we also consider an alternate model.
In this model, instead of accounting for overflows from the donor's outer Lagrangian
point, we assume part of the mass streaming through L$_1$ is ejected from L$_2$.
To do this, we compute an additional optically thin rate from Sec.~\ref{sec:thin} by replacing
$R_\mathrm{RL}$ with $R_{\mathrm{L}_2}$ and evaluating
$\mathrm{d}A/\mathrm{d}\Phi$ at L$_2$ rather than L$_1$. Whenever
$R_\mathrm{ph}>R_{\mathrm{L}_2}$, we consider the contribution to the optically thick mass
transfer rate transferred through the L$_1$ plane but above the equipotential of
L$_2$, which is given by Eq.~\eqref{equ:thick2} with a modified lower interval for
the integral:
\begin{eqnarray}
   \dot{M}_{{\rm
   thick,\;above\;L}_2}=\int_{P(R_{\mathrm{L}_2})}^{P(R_{\rm
   ph})}\left(\frac{P_0}{\rho_0}\right)^{1/2}F_3(\Gamma_1)\frac{\mathrm{d}A}{\mathrm{d}\Phi}\mathrm{d}P_0.
   \label{equ:thick3}
\end{eqnarray}
Since this expression describes a flow through L$_1$,
$\mathrm{d}A/\mathrm{d}\Phi$ has to be evaluated at L$_1$. We assume both the
optically thin component through L$_2$ and the optically thick component going through L$_1$ above
the L$_2$ equipotential are ejected from the system
taking the specific angular momentum that corresponds to the L$_2$ point,
\begin{eqnarray}
   j_{\mathrm{L}_2} =
   \frac{2\pi}{P_\mathrm{orb}}a^2\left(\frac{M_a}{M_a+M_d}-\hat{X}_{\mathrm{L}2}\right)^2.\label{equ:jout}
\end{eqnarray}
Similar to what was discussed in the previous section, material could be torqued
or form a circumbinary disk before being ejected, so we also consider extreme
cases where we increase the specific angular momentum removed by factors of $4$ and $9$.

The default setup of our simulations is the one described in Sec.~\ref{sec:mout},
but we also perform simulations with this modified model with L$_2$ outflows to test the robustness of our
results. For simplicity, in the remainder of the manuscript we drop the subscript $0$ to refer to
properties from our hydrostatic stellar models, and refer to the photospheric
radius of a star as $R$.

\subsection{Common envelope evolution}\label{sec:CE}

Whenever our MT prescription gives a MT rate exceeding
a given threshold $\dot{M}_\mathrm{high}$, we assume evolution will proceed through a CE phase.
We treat CE evolution following the standard energy prescription
\citep{TutukovYungelson1979, Webbink1984}
which equates the binding energy of ejected layers to the difference in orbital
energy product of an inspiral,
\begin{eqnarray}
   E_\mathrm{bind}=\alpha_\mathrm{CE} \Delta E_\mathrm{orb},
\end{eqnarray}
where $\alpha_\mathrm{CE}$ is a free parameter that represents the efficiency with
which the orbital energy ejects the envelope. Using subscripts $\mathrm{i}$ and $\mathrm{f}$ to represent the pre and post-CE properties
of the system, the difference in orbital energy is
\begin{eqnarray}
   \Delta E_\mathrm{orb}=-\frac{GM_\mathrm{d,f}M_\mathrm{a,f}}{2a_\mathrm{f}} +
   \frac{GM_\mathrm{d,i}M_\mathrm{a,i}}{2a_\mathrm{i}}, \label{equ:enbalance}
\end{eqnarray}
and we take $M_\mathrm{d,f}$ to be the core mass of the donor $M_\mathrm{core}$, while
ignoring accretion into the accretor such that $M_\mathrm{a,f}=M_\mathrm{a,i}$.
The binding energy depends on the value of $M_\mathrm{core}$,
and is computed by adding up the internal and gravitational potential energy of
the removed layers at the onset of CE,
\begin{eqnarray}
   E_\mathrm{bind}=\int_{M_\mathrm{core}}^{M_\mathrm{d,i}}\left(-\frac{Gm}{r}+\alpha_{\rm
   th}u\right) \mathrm{d}m,\label{equ:ebindint}
\end{eqnarray}
where $u$ is the specific internal energy of the gas and we include an
additional free parameter, $\alpha_\mathrm{th}$, which was introduced by
\citet{Han+1995} to represent the efficiency with which thermal energy can be
used to eject the envelope. In this work, we include the contribution of
recombination energy from hydrogen and helium on $u$. For simplicity, we also
assume $\alpha_\mathrm{th}=1$ throughout.

Computing the binding energy requires one to know $M_\mathrm{core}$, which is a
non-trivial problem that can lead to large variations in the predicted binding
energy (see for instance the discussion in \citealt{Ivanova+2013}). Our goal is
to determine $M_\mathrm{core}$ self consistently in our simulations by modelling
the stripping of the stellar envelope. For this, at the onset of CE we compute
the value of $E_\mathrm{bind}$ for all choices of $M_\mathrm{core}$ in the stellar
model. The CE phase is then modelled by artificially removing mass from the star
and computing at each step the binding energy one would have obtained from the
pre-CE model if $M_\mathrm{core}$ were assumed to be the current mass of the star.
In this way Eq.~\eqref{equ:enbalance} can be used to determine the final orbital
separation $a_\mathrm{f}$ as a function of $M_\mathrm{core}$, and the mass losing stellar
model can be used to determine the point at which the star would contract inside
its Roche lobe.

We do not simply force a rapid mass loss rate on
our simulations until we find $R<R_\mathrm{RL}$, but instead softly turn off the
mass loss rate as the star goes inside its Roche lobe, to determine the mass
coordinate at which the star would be undergoing contraction in the absence of
mass loss. In particular, we force a mass loss rate from CE evolution of
\begin{eqnarray}
   \begin{aligned}
      \log_{10}\dot{M}_{\rm
      CE}=\qquad\qquad\qquad\qquad\qquad\qquad\qquad\qquad\qquad\quad\\
\begin{cases} 
   \displaystyle \log_{10}\dot{M}_\mathrm{high} & \displaystyle \frac{R}{R_\mathrm{RL}} > 1\\
   \displaystyle \log_{10}\dot{M}_\mathrm{high} +  \frac{1-R/R_{\rm
   RL}}{\delta}\log_{10}\left(\frac{\dot{M}_\mathrm{low}}{\dot{M}_{\rm
   high}}\right) & \displaystyle 1-\delta < \frac{R}{R_{\rm
   RL}} < 1
\end{cases},\label{equ:cemdot}
   \end{aligned}
\end{eqnarray}
where we interpolate between a high MT rate $\dot{M}_\mathrm{high}$, meant to reproduce a
near adiabatic response, and a small MT rate, $\dot{M}_\mathrm{low}$,
which should be well below the thermal timescale of the envelope. In particular
in our simulations we choose $\dot{M}_\mathrm{high}=1\;M_\odot\;\mathrm{yr}^{-1}$,
$\dot{M}_\mathrm{low}=10^{-5}\;M_\odot\;\mathrm{yr}^{-1}$ and $\delta=0.02$. Our choice of
$\dot{M}_\mathrm{high}$ is based on the thermal timescale of
the envelope estimated as $\tau_\mathrm{th}=E_\mathrm{bind}/L$ and the associated MT
rate $\dot{M}_\mathrm{th}=M/\tau_\mathrm{th}$. For the $30M_\odot$ donor considered in this
work evolved as a single star, and assuming a core--envelope boundary at the innermost point where $X>0.1$
\citep{DewiTauris2000}, we find that $\dot{M}_\mathrm{th}$ does not exceed
$0.1\;M_\odot\;\mathrm{yr}^{-1}$. Similarly, our choice of $\dot{M}_\mathrm{low}$ is
such that its comparable to MT happening on a nuclear timescale. 
The dependence of our results on the free parameters $\dot{M}_\mathrm{high}$,
$\dot{M}_\mathrm{low}$ and $\delta$ is discussed in Appendix~\ref{app:numerics},
where we show how the outcome of some of our CE models are affected by these
choices.

If the star contracts below $R/R_\mathrm{RL}=0.98$ then we assume the CE phase
finishes, and proceed with the evolution of the model in the standard way,
applying our MT formalism described in the previous subsections.
As pointed out by \citet{Ivanova2011}, a star can contract
significantly during a phase of rapid mass loss, but re-expand by orders of
magnitude on a few thermal timescales if mass loss is suddenly shut-off. Because of this, they determine the core
mass of a stellar model by performing rapid mass loss simulations and finding at
which mass coordinate the resulting star would not re-expand after mass loss is
shut off. In our simulations we do find re-expansion after CE, which according
to \citet{Ivanova2011} would indicate the core boundary is at a deeper mass
coordinate, but find that the ensuing MT phase is stable when using our MT
prescription. 

This method is similar to the one presented by \citet{DeloyeTaam2010} and
\citet{Ge+2010}, who compute the adiabatic response of the stellar radii to
rapid mass loss and compare it to the evolution of the Roche lobe radii in a
binary to determine at which point a binary would detach. The main difference is
that our method determines the mass coordinate at which detachment would happen
absent mass loss, capturing the non-adiabatic response of the star near the end
of CE. Another similar method is the one presented by \citet{Eldridge+2017},
which is integrated into detailed binary evolution calculations but
does not consider the contribution of thermal or recombination energy.

\section{Results} \label{sec:results}

\begin{figure}
   \includegraphics[width=\columnwidth]{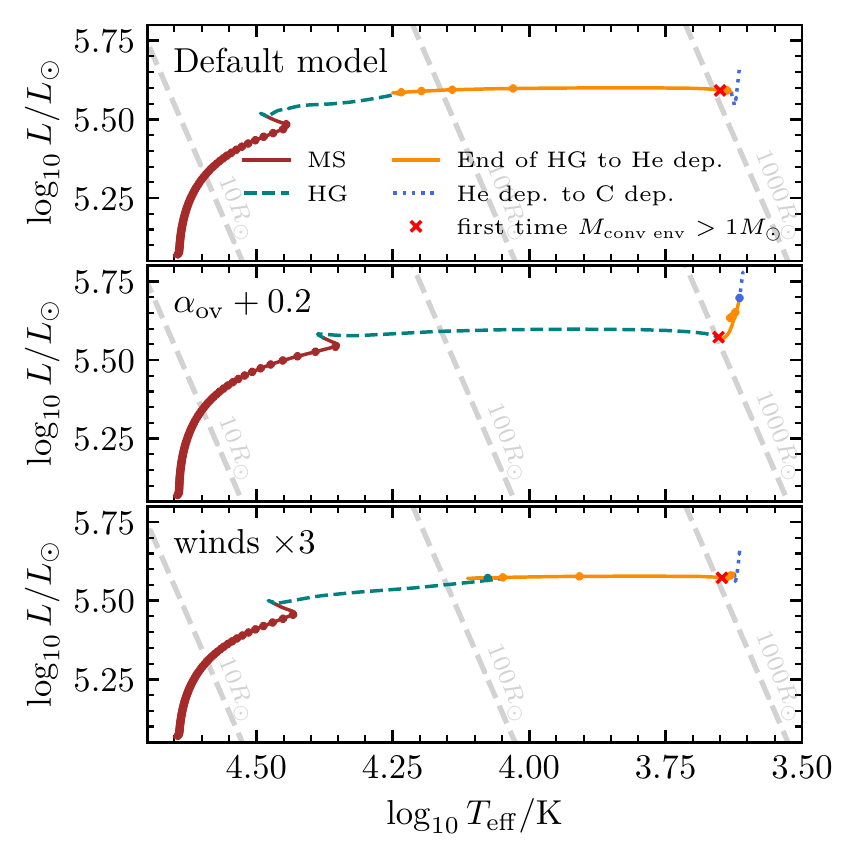}
   \caption{Evolution of a $30M_\odot$ single star with a metallicity of
   $Z_\odot/10$, separated into different evolution phases. Dots indicate equal
   time intervals of $10^5$~years. The Hertzsprung gap is defined as the
   evolutionary stage between the terminal-age
main sequence and the moment where the ratio of the
   nuclear burning luminosity and the surface luminosity {satisfies
   $|\log_{10}(L_\mathrm{nuc}/L)|<0.001$ for at least a thermal timescale
   estimated as $GM^2/RL$, or the star develops a convective envelope $>1M_\odot$}. {\textit{Top:} Evolution under our default set of
   physical inputs. \textit{Middle:} Evolution with core overshooting increased from
   $0.335$ to $0.535$ pressure scale heights. \textit{Bottom:} Evolution with wind
   mass loss rates tripled.}}\label{fig:HR}
\end{figure}

\begin{figure*}
   \begin{center}
   \includegraphics[width=1.9\columnwidth]{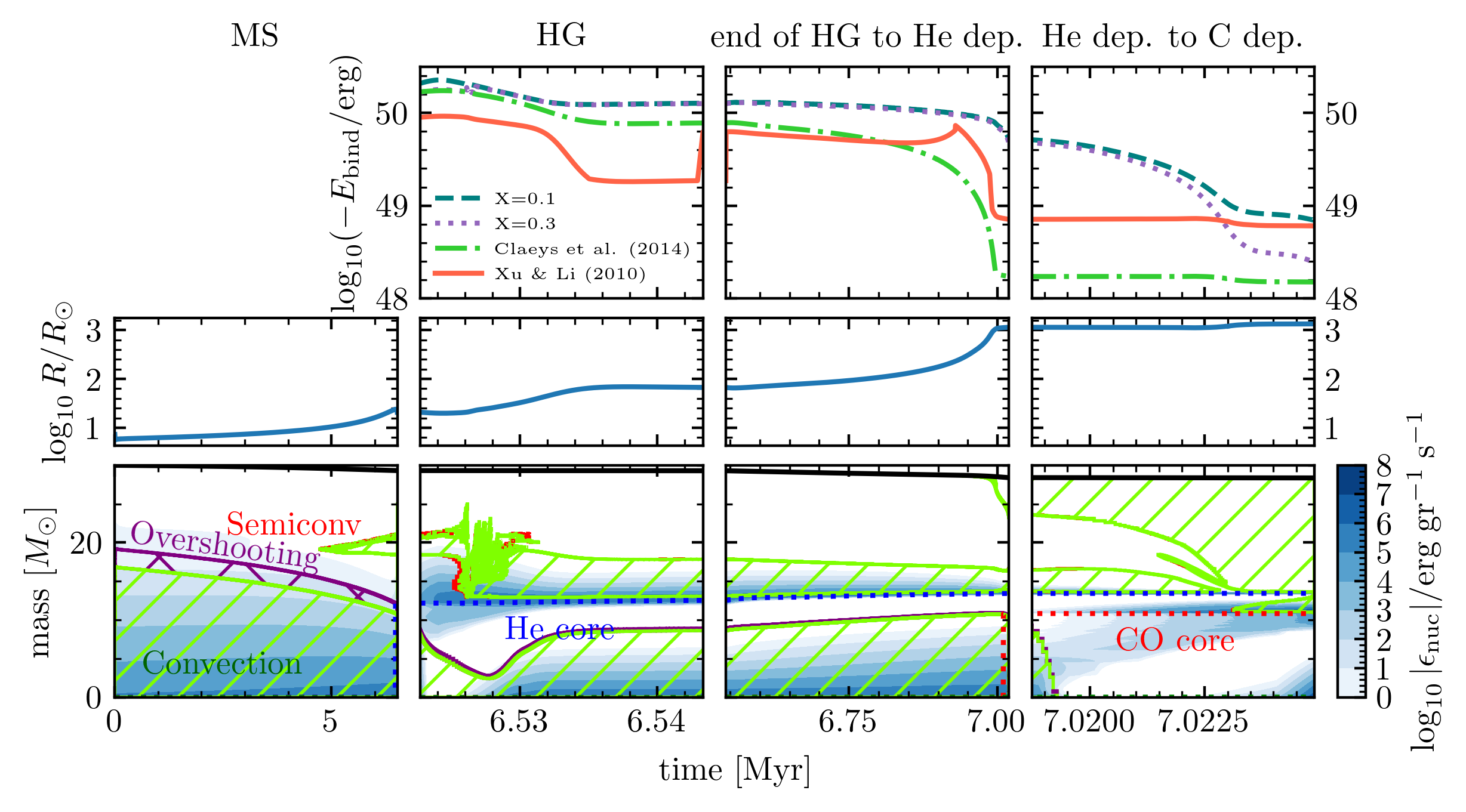}
      \vspace{-0.2in}
   \end{center}
   \caption{Kippenhahn diagram for our default $30M_\odot$
model, separating different evolutionary stages as well as showing the evolution
of the stellar radii. "He dep." and "C dep." stand for core helium and carbon
   depletion respectively. In addition, we show the result of computing $E_\mathrm{bind}$ under the assumption that the core--envelope
boundary $M_\mathrm{core}$ at which CE would finish is at the innermost mass coordinates where the
   hydrogen mass fraction is $X=0.1$ or $X=0.3$. We only show $E_\mathrm{bind}$
   after the formation of the helium core, so no information is given on the
   main sequence. To provide a comparison point to
   rapid population synthesis codes, we also show the resulting $E_\mathrm{bind}$
   from the fits of \citet{Claeys+2014} and \citet{XuLi2010}.
   }
   \label{fig:kipp}
\end{figure*}

In this work we focus on the evolution of a $30M_\odot$ star at a metallicity of
$Z_\odot/10$, in order to assess how some of the differences expected from
evolution in low metallicity environments can affect the outcome of either MT
or CE evolution. As we do not perform a broad analysis for different
donor masses and metallicities, we cannot perform population synthesis to
compute, for example, the resulting rates of binary BHs observable by ground-based detectors. However, this serves to assess if some of the assumptions
commonly used in rapid population synthesis codes are sensible. In particular,
\citet{Belczynski+2010} argue that for BH progenitors ejection of the envelope during CE evolution
is more likely to happen in low metallicity environments, as these systems are
predicted to halt their expansion in the Hertzsprung gap (HG) before rising
on the Hayashi line as red supergiants, spending a significant fraction of their
core helium burning lifetime as blue supergiants. After core helium ignition, it is argued that
a well defined core--envelope structure is formed which allows for a successful
envelope ejection. A halted expansion after the main sequence is a common
feature in stellar evolution calculations at low metallicity {(see
Appendix~\ref{app:codes})}.

Before discussing grids of binary simulations, it is instructive to study how
our $30M_\odot$ donor evolves as a single star. The top panel of
Fig.~\ref{fig:HR} shows the evolution in the Hertzsprung--Russell (HR) diagram for this star under
the physical assumptions described in Sec.~\ref{sec:methods},
for which we find that expansion is halted before becoming a red
supergiant.
Most of the helium-burning phase of this star would be
spent as a blue supergiant with a convective envelope developing only near core
helium depletion. This type of evolution is sensitive to physical
assumptions in the model. In particular, Fig.~\ref{fig:HR} shows
the effect of increasing overshooting during core hydrogen burning, as well as
increasing wind strengths. {With increased overshooting} the star undergoes rapid expansion
after the main sequence and becomes a red supergiant before core helium
ignition. {The model with a higher mass loss rate slows down its
expansion before becoming a red supergiant, but expands to a larger radius
during the HG than our default model}. These variations are not captured in rapid population synthesis codes,
in particular those reliant on the fits to single star evolution of
\citet{Hurley+2000} which are based on detailed stellar models without mass
loss and with a fixed amount of overshooting. Stellar winds may be
overestimated in our calculations \citep{Bjorklund+2020}, while observations of massive
stars suggest our choice for overshoot from a convective hydrogen-burning core
is underestimated for stars with masses $>15M_\odot$ \citep{Castro+2014}.
Keeping in mind that due to uncertainties in stellar evolution
a halted expansion in the HG is uncertain, we can still study what would be the
consequences in the evolution of a star if this effect is real.

Focusing only on our model {that uses our default set of
assumptions},
we assess if core helium ignition has a significant
impact on the core--envelope structure of the star which could favor CE ejection.
Figure~\ref{fig:kipp} shows a Kippenhahn diagram for the $30M_\odot$ model
indicating how the stellar radius and binding energy of the envelope evolve.
Assuming that the core--envelope boundary is located at the points where the
hydrogen mass fraction drops below either $X=0.1$ or $X=0.3$ (similar to  \citealt{DewiTauris2000}), we find that the
binding energy of the star remains at $\sim 10^{50}~\mathrm{erg}$ from terminal-age
main sequence (TAMS) to the point where
a convective envelope is developed. This happens despite a $\sim30$ fold increase in the
stellar radius. {This model experiences the formation and merger of multiple
intermediate convection zones during the HG phase, so we have performed
convergence tests to verify that our results are not significantly affected by
changes in temporal and spatial resolution during this phase (see Appendix~\ref{app:resolution}).}

\begin{figure}
   \includegraphics[width=\columnwidth]{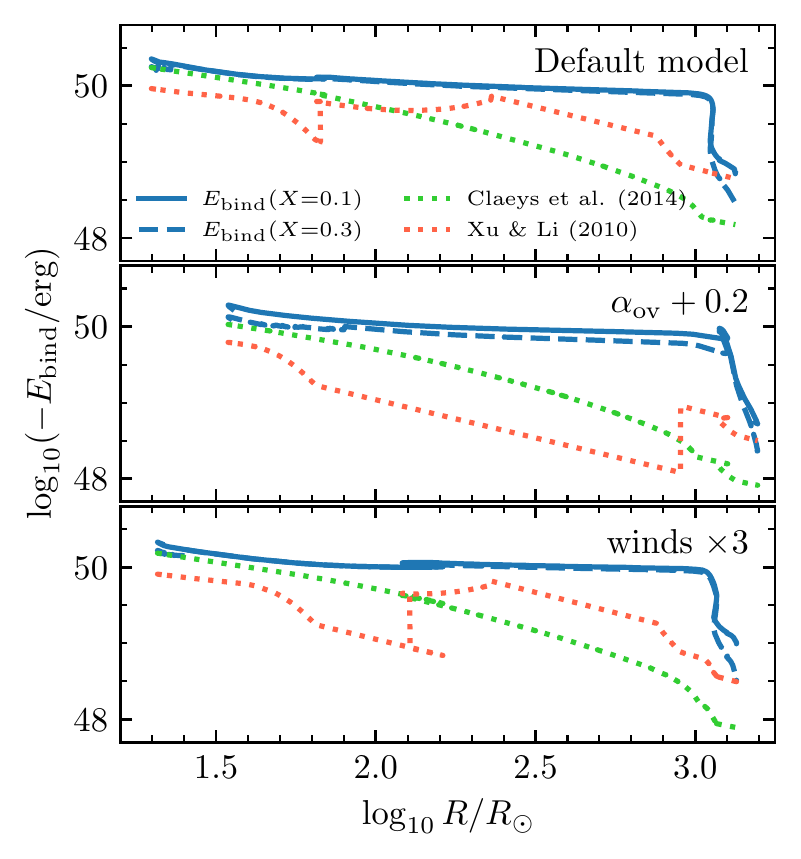}
   \caption{Binding energy of the envelope of a $30M_\odot$ post main-sequence
   star as a function of radius. Tracks are shown for our default set of
   physical assumptions, as well as for a model with increased overshooting and
   one with increased mass loss rates. For each model we also show the binding
   energy resulting from the fits of \citet{XuLi2010} and
   \citet{Claeys+2014}}\label{fig:ebind_alt}
\end{figure}

If one computes the binding energy using, for example,
the fit provided by \citet{Claeys+2014} rather than the structure of the model,
the resulting binding energy is always underestimated, with more that an order
of magnitude difference in late stages.\footnote{\citet{Claeys+2014}
provide fits for the dimensionless $\lambda$ parameter which is defined as
$E_\mathrm{bind}=G (M-M_\mathrm{core})M/\lambda R$, where $M_\mathrm{core}$ is the core mass that would remain after CE. We compute the corresponding
binding energy by assuming $M_\mathrm{core}$ is the helium core mass of our model.
{A description of how we compute these $\lambda$ fits from
our stellar models is given in Appendix~\ref{app:lambda}}} In particular, this fit to the binding
energy is used in population synthesis calculations of GW
sources, including 
the \texttt{MOBSE} \citep{GiacobboMapelli2018} and \texttt{COSMIC} \citep{Breivik+2020}
codes. Another parametrization of the binding energy that is used in published
rapid population synthesis results is that of \citet{XuLi2010}, which is an option
in the \texttt{STARTRACK} \citep{Dominik+2012} and  \texttt{COMPAS}
\citep{Neijssel+2019} codes.\footnote{The \citet{XuLi2010} fits
only cover masses up to $20M_\odot$. For higher masses \citet{Dominik+2012}
requested additional models and performed fits to these. Although the fitting
formulae are not given in \citet{Dominik+2012}, they can be found on the
publicly available source code of the \texttt{COMPAS} code at
\href{https://github.com/TeamCOMPAS/COMPAS/}{github.com/TeamCOMPAS/COMPAS/}.}
{In this case we} also find a consistent underestimation of the
binding energy {except for late stages near core-carbon burning
where the envelope becomes almost fully convective}. Both \citet{Claeys+2014} and \citet{XuLi2010} include
internal energy in their fits but not recombination energy. A
systematic underestimation of the binding energy in population synthesis
calculations implies an overestimation of post-CE orbital separations, or a
prediction of successful envelope ejection in systems that would merge during CE.
{We have also compared the binding energies of our models with different
overshooting and wind strengths to the results obtained from the
\citet{Claeys+2014} and \citet{XuLi2010} fits, and find as well that the fits
consistently underestimate the binding energy of the envelope (see
Fig.~\ref{fig:ebind_alt}). Binding
energies computed using the \citet{XuLi2010} fit experience a jump when the star
transitions out of the HG, which for the track with extra overshoot does not
happen until the star begins to rise in the Hayashi line.}

\begin{figure}
   \includegraphics[width=\columnwidth]{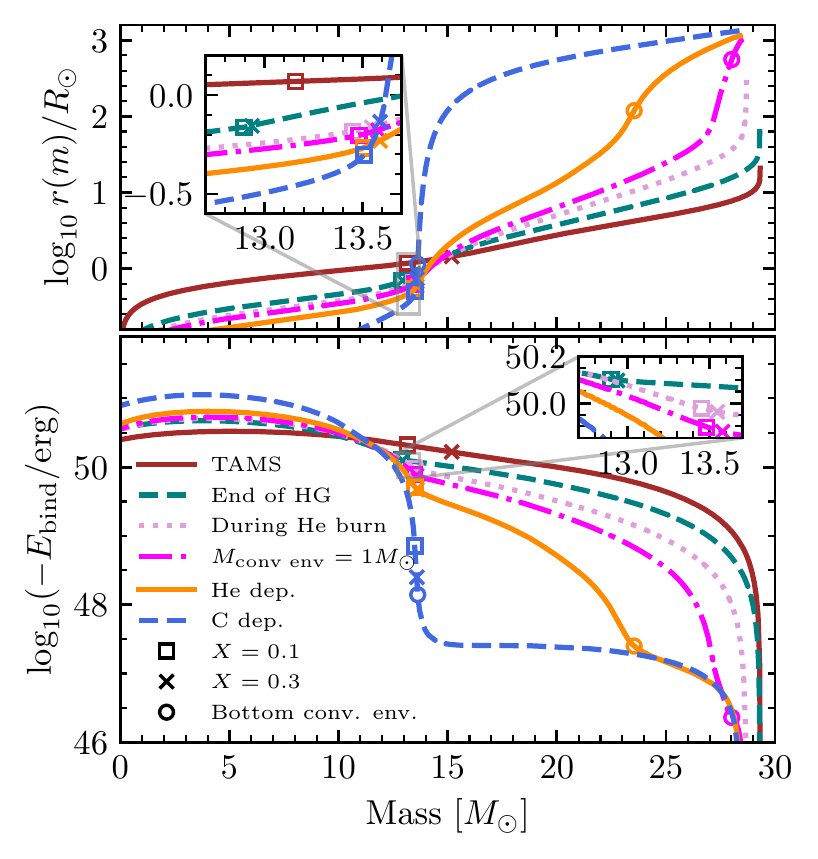}
   \caption{Radius (\textit{top}) and binding energy (\textit{bottom}) profiles for a $30M_\odot$ single star at
   different evolutionary stages. {A profile near the end of core helium-burning
   ($Y_\mathrm{c}=0.08$) but before the formation of a convective envelope is
   included}. Also, a profile is shown corresponding to the beginning of
   the formation of a deep convective envelope, taken as the first point
   in the evolution where the mass of the convective envelope $M_\mathrm{conv\;env}$ is larger than $1M_\odot$.}\label{fig:ebind}
\end{figure}

The constancy of the binding energy while the star expands with a radiative
envelope is also illustrated by Fig.~\ref{fig:ebind}, which shows
the full binding energy profiles at different evolutionary phases. In
particular, between the end of the HG and the
point where a deep convective envelope starts to develop, the radius of the
star increases by more than {an order of magnitude}, but this expansion only happens in
the outermost mass layers of the star. It is only after the formation of the
convective envelope that a large fraction of the envelope mass is pushed to
large radii with correspondingly low binding energies that would make envelope
ejection during CE feasible.
All of this points to the ignition of helium at the stellar core having
no impact on the binding energy of the envelope, and also since the ignition of
helium does not modify the mode of energy transport in the outer layers, the
stability of MT should not be impacted in a fundamental way either.

The near constancy and high value of the binding energy before the formation of a convective
envelope has also been studied comprehensively by \citet{Kruckow+2016} and \citet{Klencki+2020b},
covering a large range of masses and metallicities. However, the binding
energies computed by \citet{Kruckow+2016} and \citet{Klencki+2020b} as well as those shown in Fig.~\ref{fig:kipp} rely on an assumed boundary for the core
mass. By performing binary simulations using the method described in Sec.~\ref{sec:CE} to determine the core--envelope boundary self consistently we can
verify if these computed binding energies with an assumed boundary are a good
approximation.

\subsection{Grid of simulations}\label{sec:grids}

Using our prescription for MT and our method to calculate
the outcome of CE evolution, we computed a grid of models consisting of a
$30M_\odot$ donor with a BH accretor in a circular orbit.
Our method for CE evolution still leaves the efficiency $\alpha_\mathrm{CE}$ as a
free parameter, and we compute simulations with both $\alpha_\mathrm{CE}=1$,
{$0.4$, $0.2$} and
$0.1$. The grid spans a period range between
$-0.3<\log_{10}(P_\mathrm{i}/\mathrm{d})<3.5$ and initial mass ratios $q=M_\mathrm{BH}/M_{1,\mathrm{i}}$ between $0.01$ and $0.59$. {This includes BHs with
masses as low as $0.3M_\odot$ that are not expected to be formed through stellar
evolution, but are included to test the limits of MT stability}. This allows us to cover the entire
parameter space where CE happens, as well as to cover systems from the boundary
of RLOF at ZAMS, to effectively single star evolution at
large periods. Owing to the small amount of radial expansion happening in the
phases where the star has a convective envelope, we adopt two different
resolutions in period for our grid. For $\log_{10}(P_\mathrm{i}/\mathrm{d})<3.2$ we use a
spacing of $\Delta \log_{10}(P_\mathrm{i}/\mathrm{d})=0.1$, while at higher initial orbital
periods we increase the resolution in period to $\Delta \log_{10}(P_\mathrm{i}/\mathrm{d})=0.01$.
The resolution in mass ratio of the grid is $\Delta q=0.02$.

\begin{figure}
   \includegraphics[width=\columnwidth]{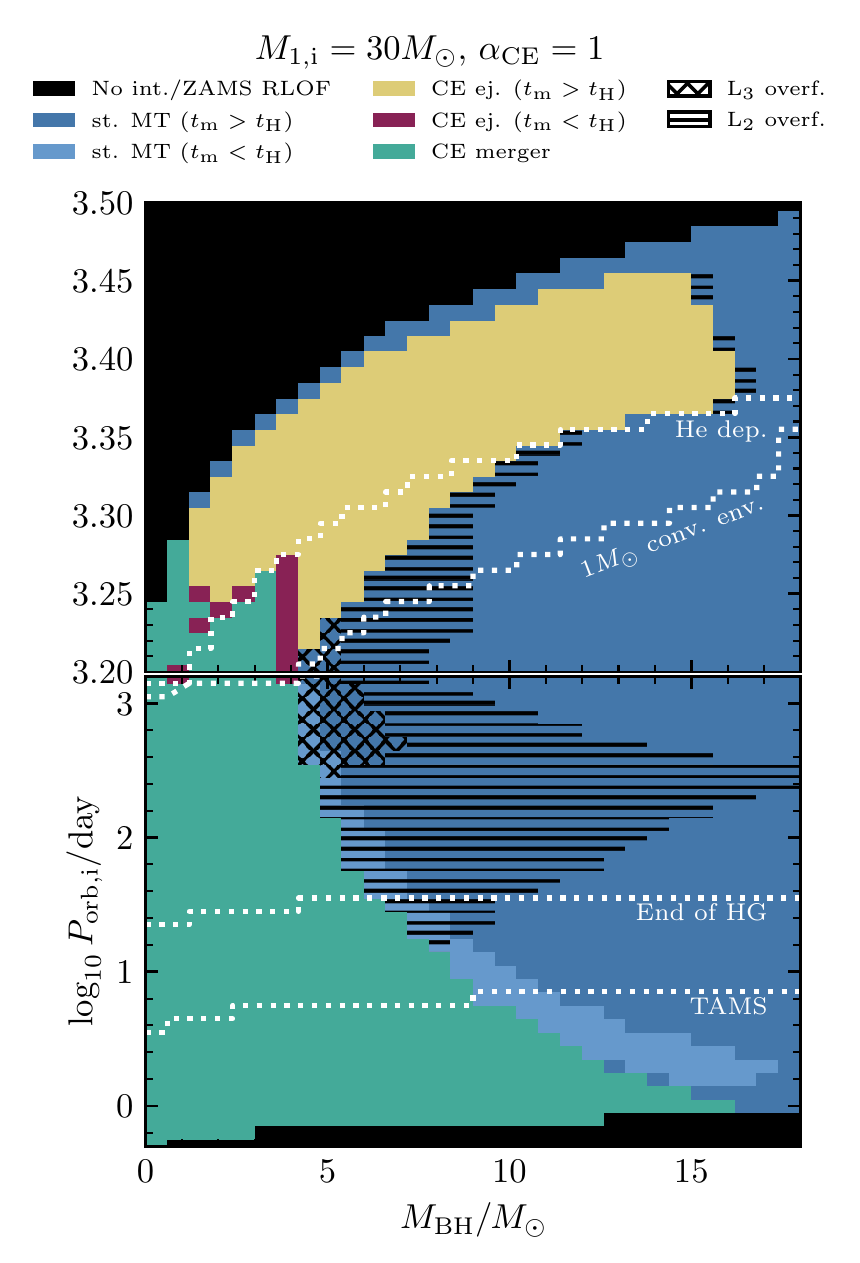}
   \caption{Summary of final outcomes for simulations of circular binaries
   consisting of a BH and a $30M_\odot$ star at a metallicity of $Z_\odot/10$,
   and a CE efficiency parameter $\alpha_\mathrm{CE}=1$.
   Horizontal dotted lines indicate boundaries for interaction before different
   evolutionary stages of the star. For systems that undergo stable MT,
   {black horizontal lines indicate regions where the donor excedeed
   the L$_2$ equipotential, while cross hatched regions mark regions where also
   the L$_3$ equipotential is exceeded}. Black
   rectangles indicate systems that do not interact or would result in a Roche
   lobe filling system at ZAMS. Systems that undergo stable MT or eject their
   envelope during CE evolution are denoted by "st. MT" and "CE ej."
   respectively, and separated into systems forming binary BHs that would merge
   in less or more than a Hubble time. Systems marked as "CE merger" merge
   during the CE phase.}\label{fig:grid1}
\end{figure}

\begin{figure*}
   \includegraphics[width=2\columnwidth]{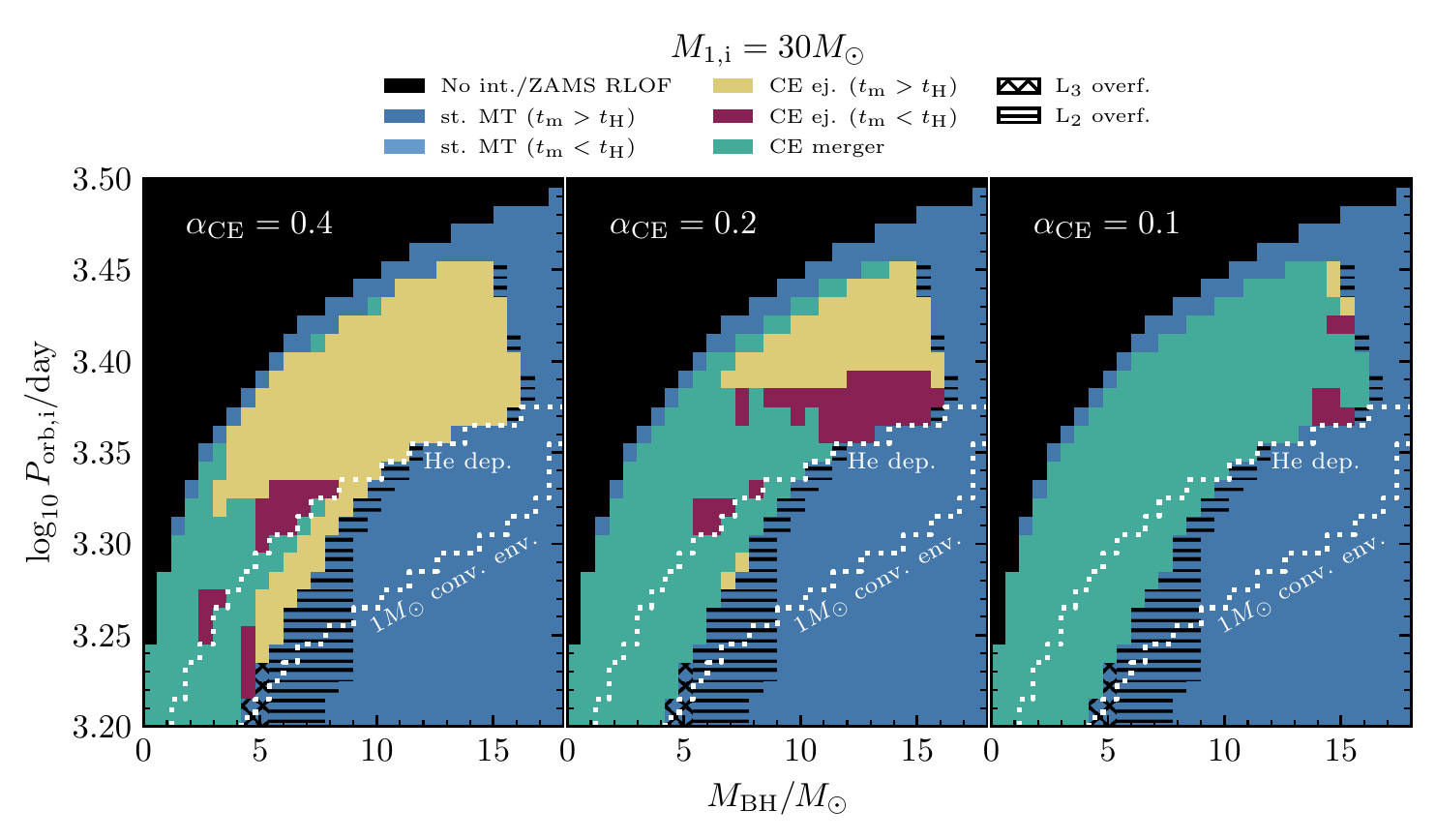}
   \caption{Same as Fig.~\ref{fig:grid1} but for CE efficiency
   parameters of $\alpha_\mathrm{CE}=0.4$, $0.2$ and $0.1$.}\label{fig:grid2}
\end{figure*}

The final outcomes of our grid with $\alpha_\mathrm{CE}=1$ are shown in Fig.~\ref{fig:grid1}. As a simple approximation to determine if a simulation can
produce a merging binary BH, for stars that reach core carbon depletion
we assume the donor star collapses directly to form
a BH with a mass equal to its baryonic mass. Our $30M_\odot$ donor produces
final helium core masses $\sim 14M_\odot$, for which the formation of a BH
through direct collapse is indeed expected \citep{Fryer1999}. From our simulations we do not find
any case of successful envelope ejection from CE evolution for initial orbital periods
below $\log_{10}(P_\mathrm{i}/\mathrm{d})<3.2$. Below this period range the boundary between
systems undergoing stable MT and those merging during CE evolution
varies continuously with period, going from $q\sim 0.5$ for systems interacting
close to ZAMS, down to $q\sim 0.15$ for cases where interaction happens
right before the formation of a deep convective envelope. The
boundary is not sensitive to whether interaction happens before or after the TAMS,
or before or after the end of the HG. This is to be expected, as the radial
response to rapid mass loss is dependent on the mode of energy
transport in the stellar envelope \citep{HjellmingWebbink1987},
and the envelope of the star across all these phases is radiative.
{\citet{Pavlovskii+2017} also find stable MT between a star and a BH for such
extreme mass ratios for different donor masses, but owing to their coarse mass
ratio sampling (no more than three different mass ratios for each donor mass)
cannot describe in detail how the boundary for stability shifts with initial
orbital period.}

The lack of systems surviving CE evolution between the end of the HG and the
formation of a deep convective envelope can be understood in terms of the
binding energy of the envelope. As was discussed for the
single $30M_\odot$ model, the binding energy during this phase is $\sim
10^{50}~\mathrm{erg}$. Considering evolution between the end of the HG and the
formation of a convective envelope, we find that CE evolution only happens for BH masses
$\lesssim 5 M_\odot$, and that stripping the hydrogen envelope of the $30M_\odot$
star would result in a $\sim 14 M_\odot$ helium core. For these masses, and
assuming a fully efficient use of the orbital energy to eject the
envelope, in order to release $10^{50}~\mathrm{erg}$ the orbital separation needs to be reduced
below $1.3R_\odot$, an orbit for which a helium star of $14 M_\odot$
would overfill its Roche lobe.
Population synthesis calculations that underestimate the binding
energy during this phase and find successful CE ejection, for example, by using fits such as those of
\citet{Claeys+2014} or by assuming a fixed value of the $\lambda$ parameter,
potentially overestimate the rate of
formation of merging binary BHs through CE evolution
(Gallegos-Garcia et al., in preparation).

The boundary for unstable MT is also right at the place where it
allows for a narrow band of systems undergoing stable MT that produce
a binary BH merging in less than a Hubble time. In this case, the shrinkage of
the orbit from stable MT is sufficient to produce a compact BH
binary \citep{vandenHeuvel+2017}, and we find this outcome for a wide range of
initial orbital periods ranging from $1$ to $1000$~days. Multiple systems undergoing
stable MT fill their outer Lagrangian point during that process, but remain stable despite the
increased angular momentum loss. For wider orbital periods,
where interaction happens after the development of a deep convective envelope,
the critical mass ratio for stability shifts upwards (up to $q\sim0.5$ for
systems interacting after core helium depletion). The majority of our models with an assumed
efficiency of $\alpha_\mathrm{CE}=1$ that manage to eject their envelopes during CE
evolution result in wide BH binaries, for which GW radiation is
insufficient to produce a merger in less than a Hubble time. Figure~\ref{fig:grid2} shows the outcome of our simulations with CE
efficiencies of {$\alpha_\mathrm{CE}=0.4$, $0.2$ and $0.1$, in which case we find that most systems
that produce wide binary BHs at $\alpha_\mathrm{CE}=1$ would merge during CE
evolution rather than produce merging binary BHs at lower efficiencies.}

\begin{figure}
   \includegraphics[width=\columnwidth]{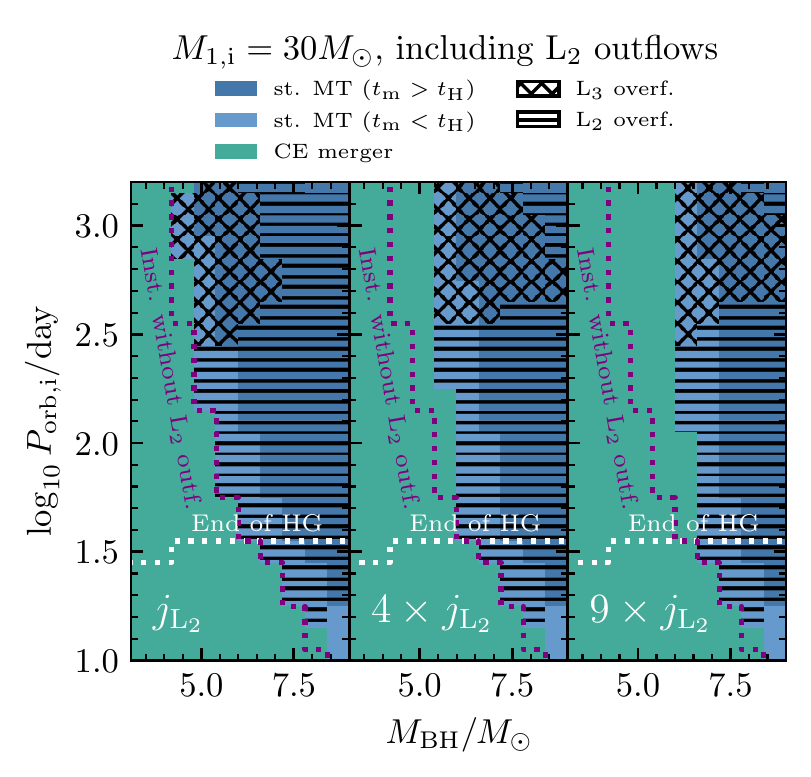}
   \caption{Same as Fig.~\ref{fig:grid1} but for a subset of models
   computed using the method described in Sec.~\ref{sec:L2out} that considers
   outflows from L$_2$ rather than L$_3$ when $M_{\rm a}<M_{\rm d}$. Each panel
   indicates simulations performed with different specific angular momentum
   assumed for the material outflowing through L$_2$, which is taken to be
   $j_{{\rm L}_2}$, $4j_{{\rm L}_2}$ or $9j_{{\rm L}_2}$. The purple dotted line
   indicates the boundary between stable MT and CE evolution as determined by
   our models that account only for outflows from the outer Lagrangian point of
   the donor.}\label{fig:grid3}
\end{figure}

One important concern in our simulations that form merging binary BHs through
stable MT is that an important fraction of these undergo overflow of L$_2$ or
even L$_3$. The methods we have developed to account for this are necessarily an
approximation, as they rely on hydrostatic one-dimensional stellar models.
Moreover, the simulations shown in Fig~\ref{fig:grid1} do not account for
potential outflows through
L$_2$ when $M_{\rm a}<M_{\rm d}$, which could remove significant mass and
angular momentum from the system.
Using the alternate method described in Sec.~\ref{sec:L2out} for L$_2$
outflows we can test if the systems that form merging
binary BHs through stable MT are sensitive to the potential presence of such outflows.
Fig~\ref{fig:grid3} shows simulations done using this method that cover the
region where we find stable MT leads to merging binary BHs, including cases where
we assume the specific angular momentum removed by material outflowing L$_2$ is
up to 9 times the specific angular momentum of the L$_2$ point. Compared to our
simulations that only account for overflow of the outer Lagrangian point of the
donor, we find that the boundary for instability is pushed to higher BH masses.
However, we still find systems across a large range of periods that form merging
binary BHs, as additional angular momentum losses lead to the formation of more
compact binaries post-MT. Given this, the conclusion that stable MT can lead to
the formation of merging binary BHs across a large range of periods appears to
be robust.

As our models do not include the evolution prior to the formation of the
first BH, we cannot assess the relative likelihood of each of our individual
simulations to occur in nature. However, if we make the simplistic assumption
that after BH formation the orbital period is distributed flat in $\log_{10}P$
and the mass ratio distribution is flat in $q$, our grid with $\alpha_\mathrm{CE}=1$ gives a ratio between the number of merging binary BHs formed through
CE evolution to those formed via stable MT of {$0.017$}. The actual outcome of our
simulations for systems surviving CE evolution should be considered with care,
since we approximate a complex three-dimensional phenomenon with one-dimensional hydrostatic models. However,
even under the extreme that all our systems that survive CE evolution would
produce a binary BH that merges in less than a Hubble time, we find that the
ratio between merging binary BHs formed through CE evolution to those formed via
stable MT is {$0.35$}. This resembles the results from population synthesis
calculations of \citet{Neijssel+2019}, who find that depending on the metallicity
distribution of the star formation history most merging binary BHs that will be
observed by current generation GW detectors at design sensitivity would have
been formed through stable MT. Our results suggest that the yield of
merging binary BHs from CE evolution is overestimated at low
metallicities, which would further increase the relative contribution of stable
MT relative to CE evolution in population synthesis calculations such as those
of \citet{Neijssel+2019}.

\subsection{Mass transfer and outer Lagrangian point overflow}\label{sec:lout_outflow}

\begin{figure}
   \includegraphics[width=\columnwidth]{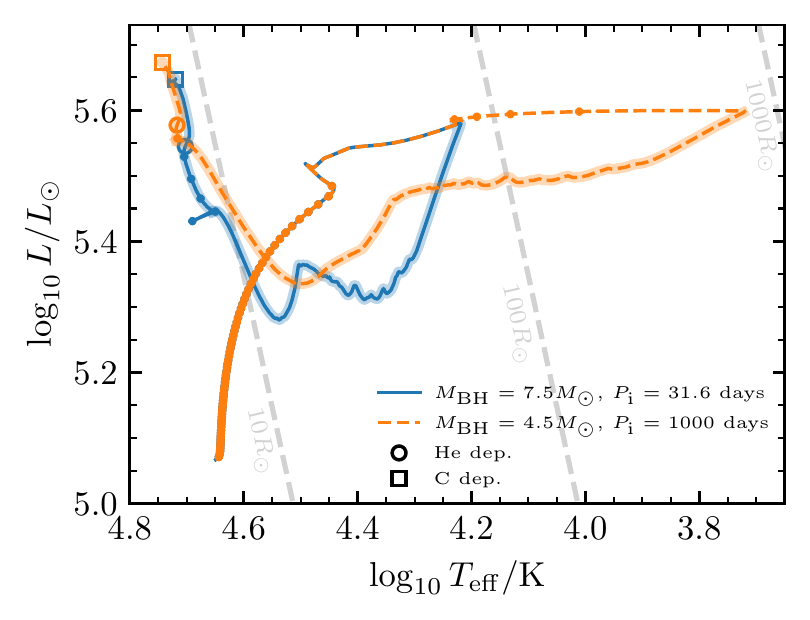}
   \caption{Evolution in the HR diagram for a $30M_\odot$ donor star in two systems
   leading to the formation of a merging binary BH through stable MT. Dots in
   the tracks indicate intervals of $10^5$~years, while thicker lines indicate
   periods of RLOF.}\label{fig:HR_MT}
\end{figure}

We consider here two cases from our simulation grid that lead to the formation of a
merging binary BH: a system interacting before the end of HG
{($M_\mathrm{BH}=7.5
M_\odot$, $P_\mathrm{i}=31.6$~days)} and one interacting after it ($M_\mathrm{BH}=4.5
M_\odot$, $P_\mathrm{i}=1000$~days). The
short period system evolves without overflowing its outer Lagrangian point,
while the long period one is expected to overflow it, so we can use it to study
how the resulting outcome is modified when accounting for this on our MT
prescription. To evaluate the impact of the changes we have made to the MT
prescription of \citepalias{KolbRitter1990}, we use four different setups for our
simulations:
\begin{enumerate}
\item A full implementation of our MT prescription
   including overflows from the outer Lagrangian point {as described
      in Sec.~\ref{sec:mout}}.
   \item Same as 1 but
without the assumption $P/\rho=kT/\mu m_\mathrm{H}$ made in \citetalias{KolbRitter1990}
(see the discussion in Sec.~\ref{sec:thick})
\item Same as 1 but but
taking the area term $\mathrm{d}A/\mathrm{d}\Phi$ to be constant, ignoring the term linear in
$\Delta \Phi$ of Eq.~\eqref{equ:dadphi_f}.
\item Using the prescription
of \citetalias{KolbRitter1990} without modification and without outflows from the
outer Lagrangian point. 
\end{enumerate}

Figure~\ref{fig:HR_MT} shows the evolution in the HR diagram for both the
{$P_\mathrm{i}=31.6$~day} and $P_\mathrm{i}=1000$~day system using our updated MT
prescription. In both cases MT first proceeds rapidly, stripping the
majority of the hydrogen envelope in a few thousand years, followed by a longer
phase ($\sim 10^5~\mathrm{yr}$) driven by the evolution of the core during core-helium
burning. In both cases, MT lowers the mass of the donor to {$\sim 13
M_\odot$}
{and carbon is depleted while still undergoing RLOF}.
The {$31.6$~day} system depletes carbon while still retaining hydrogen in
its surface (surface mass fraction of {$X_\mathrm{s}=0.40$}), with a radius of
{$8.1 R_\odot$} and an orbital
period of {$1.8$~days}. {After it first interacts the $1000$~day system 
remains always near or at RLOF}, reaching carbon depletion with a surface hydrogen mass
fraction {$X_\mathrm{s}=0.36$}, a radius of $7.5 R_\odot$ and an orbital period of
{$1.5$~days}. The
resulting binary BHs have large delay times for merger due to GW radiation, with
the {$31.6$}~day system expected to merge after
{$6.4~\mathrm{Gyr}$} and the $1000$~day
systems after {$6.0~\mathrm{Gyr}$}.

\begin{figure}
   \includegraphics[width=\columnwidth]{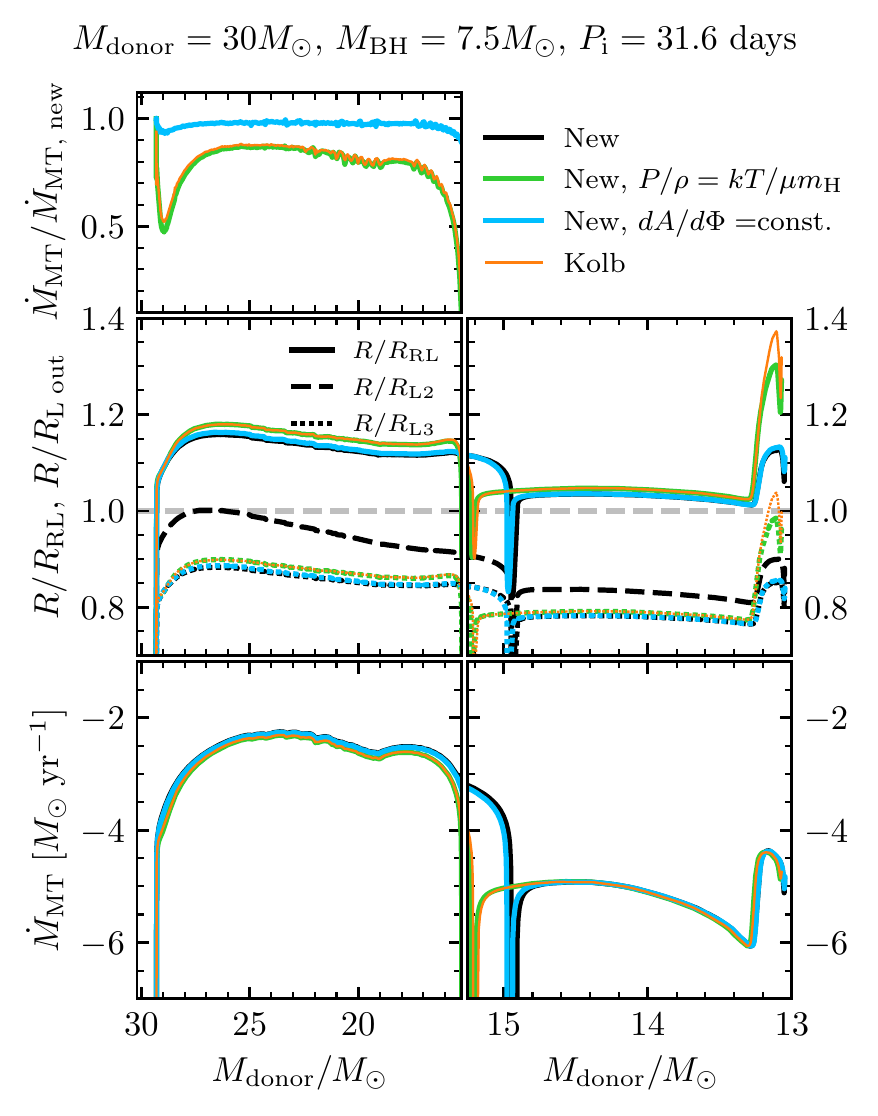}
   \caption{Evolution of the MT rate and overflow for a system with a
   $30 M_\odot$ donor star, a {$7.5 M_\odot$} BH and an initial orbital period of
   {$31.6$~days}. \textit{Top:} ratio of the computed MT rate under different
   assumptions, compared to the result from our full MT prescription. \textit{Middle:}
   Ratio of the stellar radius to the Roche lobe radius and to the volume
   equivalent radius of the {L$_2$ and L$_3$ equipotentials (for the L$_2$
   equipotential we only include the results of our default model for ease of
   visualization)}. \textit{Bottom:}
   MT rate.}\label{fig:mt1}
\end{figure}

Focusing on the {$31.6$~day} system, which we do not predict to overflow its outer
Lagrangian point, Fig.~\ref{fig:mt1} shows the evolution of
the MT rate and the Roche filling factor with the variations to the MT
prescription described before. For our updated prescription, and in the absence
of outflows from the outer Lagrangian point, the computed MT rate is
expected to be higher than that of \citetalias{KolbRitter1990}. This is because dropping the assumption of
$P/\rho=kT/\mu m_\mathrm{H}$ of \citetalias{KolbRitter1990} and accounting for the
linear term in $\mathrm{d}A/\mathrm{d}\Phi$ increases the value of the integrand in Eq.~\eqref{equ:thick2}.
{As shown in Fig.~\ref{fig:mt1}, we do find both of these assumptions
result in a lower MT rate and a higher amount of RLOF compared to our full MT
scheme. This is very noticeable} in the slower second MT phase, where our new MT
prescription predicts a maximum Roche filling factor of {$R/R_{\rm
RL}=1.16$} while the standard prescription from \citetalias{KolbRitter1990} reaches
{$R/R_{\rm
RL}=1.37$} during this phase. For this example system, we find that the largest
difference from the \citetalias{KolbRitter1990} prescription comes from dropping the
assumption $P/\rho=kT/\mu m_\mathrm{H}$, while the contribution of the
linear term in $\mathrm{d}A/\mathrm{d}\Phi$ has a much smaller impact.
Without our updated MT prescription, the \citetalias{KolbRitter1990}
prescription predicts the system overflows its outer Lagrangian point.

\begin{figure}
   \includegraphics[width=\columnwidth]{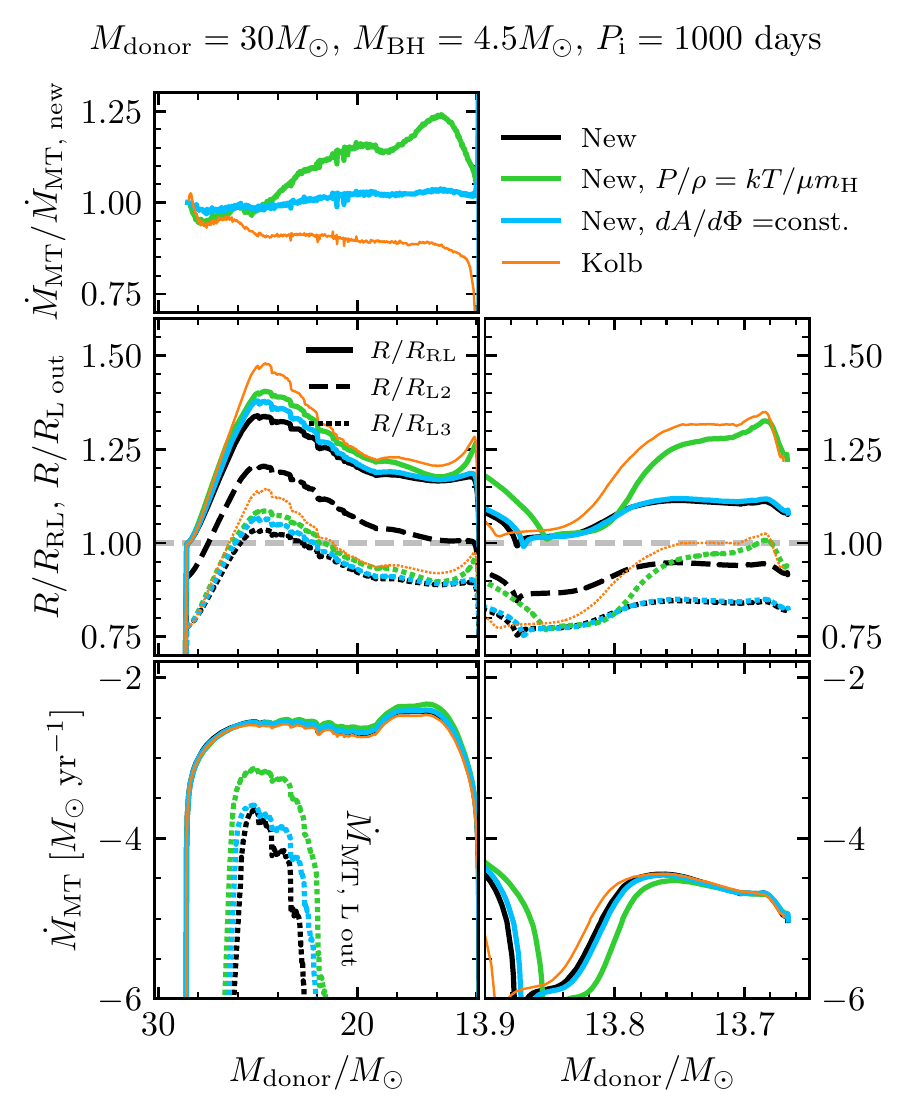}
   \caption{Same as Fig.~\ref{fig:mt1} but for a system with a $30M_\odot$
   star with a $4.5M_\odot$ BH companion at an initial orbital period of $1000$
   days.}\label{fig:mt2}
\end{figure}

The case for the $1000$~day system, which undergoes overflow of its outer
Lagrangian point in our model, is shown in Fig.~\ref{fig:mt2}. In this case we
also find that our updated prescription keeps the star significantly more
compact, with the \citetalias{KolbRitter1990} prescription reaching a Roche filling
factor of {$R/R_\mathrm{RL}=1.48$}, while our model reaches {$R/R_{\rm
RL}=1.34$}. {With the updated MT scheme the donor reaches}
an overflow of the outer Lagrangian point of {$R/{R_{\rm
L\;out}}=1.04$}, resulting in an outflow from L$_3$ that at its maximum accounts
for less than {$9\%$} of the total mass loss rate of the star. Since most mass is
still transferred through L$_1$, the additional angular momentum loss from mass
loss through L$_3$ only produces a small change in the final orbital separation,
with the model using the prescription of \citetalias{KolbRitter1990} resulting in a
final orbital period of $1.6$~days compared to the {$1.5$~day} result
when using our updated MT prescription. For this binary system we find that
during overflow of the outer Lagrangian point
including the linear term in $\mathrm{d}A/\mathrm{d}\Phi$ has an effect comparable to that of
assuming $P/\rho=kT/\mu m_\mathrm{H}$. As discussed in Appendix~\ref{app:dadphi} it
is indeed expected that the linear term in $\mathrm{d}A/\mathrm{d}\Phi$ has a larger impact for
mass outflows through L$_3$ than from L$_1$.

As the specific angular momentum that
would be carried by material being ejected through the outer Lagrangian point is
uncertain (we assume by default that it corresponds to the location of the
Lagrangian point), we also repeated the $1000$~day simulation but increased the angular
momentum loss rate associated to this effect by factors of $4$ and $9$. In both
cases we still find the system to be stable, but result in much shorter final
periods of {$1.2$~day and $0.83$~day}. This corresponds to delay
times of {$3.6~\mathrm{Gyr}$
and $1.2~\mathrm{Gyr}$, compared to the $6.0~\mathrm{Gyr}$} delay time for our default model. This
additional uncertainty could then allow for stable MT to produce
systems with short delay times, where tidal spin-up can result in a high spin
second formed BH \citep[cf.][]{Bavera+2020}.
{Another potential source of mass and angular momentum loss is the ocurrence of
L$_2$ outflows when $M_{\rm a}<M_{\rm d}$ which our default model does not account for. For
the $100$~day system we find that the donor slightly overfills the L$_2$
equipotential, exceeding the L$_2$ equivalent radius by
$0.2\%$. This small amount of overflow is not expected to play a significant
role, and indeed in our experiments with L$_2$ outflows shown in
Fig.~\ref{fig:grid3} we do not see an important difference. For the $1000$~day simulation,
however, our default model overflows the L$_2$ equivalent radius by $21\%$. Due
to this, in the simulations with L$_2$ outflows shown in Fig.~\ref{fig:grid3}
we find that the outcome is sensitive to potential outflows through L$_2$,
resulting in CE evolution and a merger if the specific angular momentum of material ejected
through L$_2$ is large.}

In summary, we find that our modifications to the MT prescriptions
of \citetalias{Ritter1988} and \citetalias{KolbRitter1990} do not have a
significant effect on the MT rate or the final masses after MT
concludes. The changes, however, do affect the size of the donor through the MT
phase, resulting in more compact stars with less overflow during MT. Systems
where we predict overflow of the outer Lagrangian point result in smaller
orbital periods than when using the prescriptions of \citetalias{Ritter1988} and
\citetalias{KolbRitter1990}, but we do not find that the additional source of
angular momentum loss has a significant impact on the stability of these
systems.

\subsection{Common envelope in donors with convective
envelopes}\label{sec:ce_examples}

\begin{figure}
   \includegraphics[width=\columnwidth]{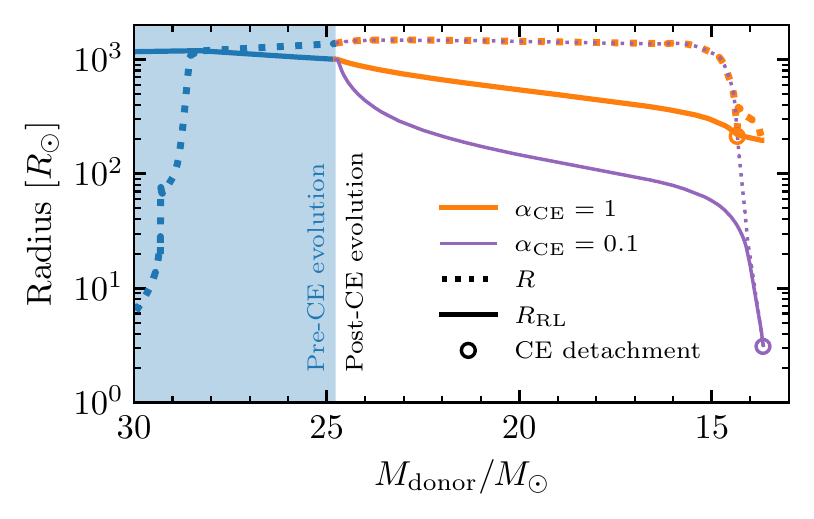}
   \caption{Evolution of the radius and Roche lobe radius for two models
   undergoing successful envelope ejection from CE evolution. Both models
   correspond to a simulation of a $30M_\odot$ star with a
   {$14.1M_\odot$} BH with
   an initial orbital period of {$2344$~days}, with the only difference between
   the simulations being the efficiency of CE, $\alpha_\mathrm{CE}=1$
   or $0.1$. The shaded blue region indicates pre-CE evolution which is identical
   for both cases. Empty circles denote for each case the moment our algorithm
   determines CE evolution would end as the star contracts within its Roche lobe
   in the absence of mass loss from CE evolution.}\label{fig:ce_example}
\end{figure}

To illustrate the result of our simulations that survive CE evolution,
we consider a particular model that results in successful ejection in both our
$\alpha_\mathrm{CE}=1$ and $\alpha_\mathrm{CE}=0.1$ grids, which has a $30M_\odot$ star with
a {$14.1M_\odot$ BH in an initial orbit of $2344$ days}. This model undergoes RLOF after
core helium depletion and the evolution of their stellar and Roche lobe radii
is shown in Fig.~\ref{fig:ce_example} for both CE efficiencies considered. Before the onset of RLOF
wind mass loss reduces the mass of the donor to {$28.2M_\odot$} and before hitting
the threshold value of $\dot{M}_\mathrm{high}=1\;M_\odot\;\mathrm{yr}^{-1}$ at which we start modeling
the system as a CE, the mass of the donor is reduced to $24.7M_\odot$. The onset
of CE happens after the star slightly overflows its outer Lagrangian point
({$R/R_{L\;\mathrm{out}}=1.08$}) at which point our MT prescription predicts
{53\%} of the mass being lost by the star goes through L$_3$. During the CE
calculation we artificially remove mass and adjust the separation according to
the binding energy of the removed layers, resulting in a sharper decrease with
mass of the Roche lobe radius for the lower CE efficiency. The stellar radius
remains larger than $10^3R_\odot$ until the mass of the star is lowered below
$15M_\odot$, at which point the radius begins to decrease sharply with
mass. The point at which we find detachment without any enforced mass loss from
CE, see Eq.~\eqref{equ:cemdot}, varies with efficiency, with the
$\alpha_\mathrm{CE}=1$ simulation detaching at a mass of {$14.3
M_\odot$} and the
$\alpha_\mathrm{CE}=0.1$ model detaching at {$13.7 M_\odot$}.

\begin{figure}
   \includegraphics[width=\columnwidth]{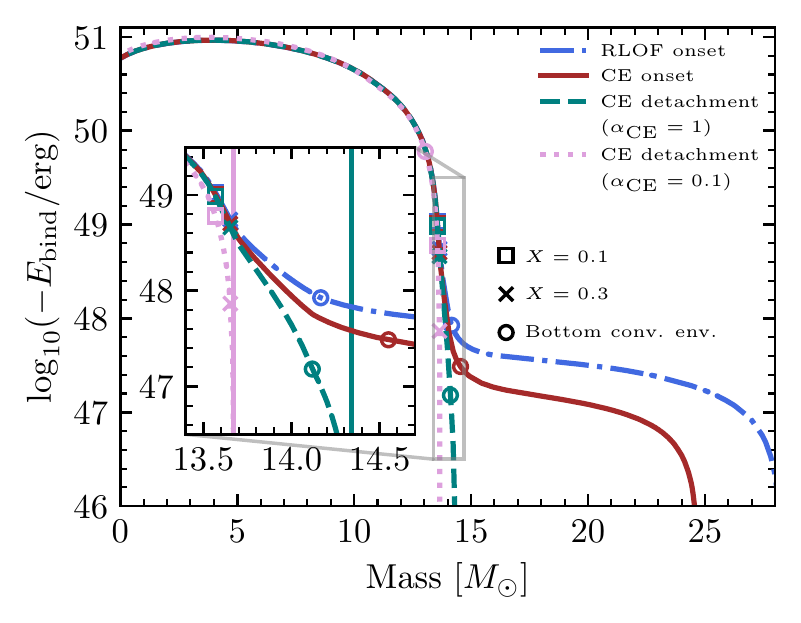}
   \caption{Binding energy profiles of the donor in a system with an initial mass of
   $30M_\odot$ and a BH companion of {$14.1M_\odot$} at an orbital period of
   {$2344$~days}. Profiles are shown at RLOF, at the onset of CE evolution, and
   after detachment from CE for $\alpha_\mathrm{CE}=1$ and $0.1$. Vertical lines
   indicate the final mass after CE evolution at the two efficiencies
   considered.}\label{fig:ebindce}
\end{figure}

Figure~\ref{fig:ebindce} shows the binding
energy profiles of the star at different phases of evolution. Between the
beginning of RLOF and the onset of CE evolution, the donor loses $\sim 3M_\odot$
of mass which modifies the binding energy of the convective envelope but not of
the deeper layers.
The binding energy of the ejected envelope changes significantly between the
$\alpha_\mathrm{CE}=1$ and $\alpha_\mathrm{CE}=0.1$ simulations with {$E_{\rm
bind}=-3.9\times 10^{47}~\mathrm{erg}$} and {$E_\mathrm{bind}=-4.3\times
10^{48}~\mathrm{erg}$}
respectively. For the fully efficient simulation all ejected mass formed part of
the convective envelope at the onset of CE. This factor of {$\sim
10$} difference on the binding energy, coupled
with the factor of $10$ difference in CE efficiency leads to very different outcomes
post-CE. The $\alpha_\mathrm{CE}=1$ simulation terminates CE evolution with an
orbital separation of $570R_\odot$, while the $\alpha_\mathrm{CE}=0.1$ simulation
finishes CE with an orbital separation of {$8.4R_\odot$}. As discussed by
\citet{DeloyeTaam2010}, the core--envelope boundary is not uniquely defined,
resulting in different values depending on the mass ratio and efficiency of the
process. One caveat of our simulations is that the amount of mass
loss previous to CE evolution is sensitive to our choice of $\dot{M}_\mathrm{high}$, which modifies the binding energy of the outer layers of the star (see
Appendix~\ref{app:numerics}). This modifies the post-CE separation for systems
we predict to remain wide, such as the $\alpha=1$ system discussed here.
Accurate predictions for these wide post-CE models requires a better description of how the
star transitions from stable MT to CE evolution.

\begin{figure}
   \includegraphics[width=\columnwidth]{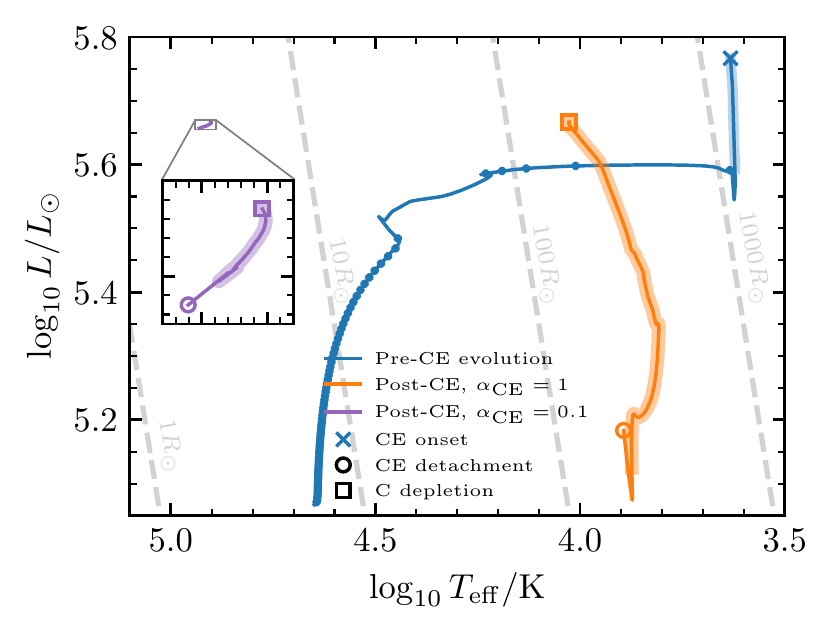}
   \caption{HR diagram showing the evolution of a system undergoing CE evolution composed of a
   $30M_\odot$ donor with a {$14.1 M_\odot$} BH companion at an initial period of
   {$2344$~days}. Results are shown for two values of the efficiency of CE
   evolution, $\alpha_\mathrm{CE}=1$ and $\alpha_\mathrm{CE}=0.1$, with the evolution
   prior to CE being identical.}\label{fig:HR_CE}
\end{figure}

After detachment, for both efficiencies we find that the donor undergoes a thermal
pulse and re-expands, filling its Roche lobe but undergoing stable MT rather than developing a CE again. Figure~\ref{fig:HR_CE} shows the
evolution in the HR diagram both before and after CE evolution. For the case of
fully efficient CE evolution, during this
additional phase of MT the star would appear as an extended
blue/yellow supergiant with a surface hydrogen mass fraction of $X_\mathrm{s}\sim0.4$. The
duration of this MT phase for the case of $\alpha_\mathrm{CE}=1$ is
$\sim 2000$~years, during which mass is transferred at a rate above the
Eddington limit of the {$14.1 M_\odot$} BH companion. For the $\alpha_\mathrm{CE}=0.1$
case the donor star has a radius of just $2R_\odot$ with a surface hydrogen mass
fraction of $X_\mathrm{s}\sim 0.3$, and transfers mass at super-Eddington rates for
{$\sim
800$~years}. Given that the Eddington luminosity of a {$14.1M_\odot$} BH is $\sim
2\times 10^{39}\;\mathrm{erg\;s}^{-1}$, during this final MT phase the system can
potentially appear as an ultra-luminous X-ray source (see \citealt{Kaaret+2017}
for a recent review), but owing to the short time before core carbon depletion,
the number of observable sources could be negligible compared to those formed by
other evolutionary pathways.

Not all our simulations that survive CE evolution have 
short lifetimes as X-ray sources. For the system we have discussed so far, this
is the case because CE evolution happens after core helium depletion. If we
consider a system with a $7.5M_\odot$ BH at an initial orbital period of $2000$~days and $\alpha_\mathrm{CE}=1$, we find that the onset of CE happens when the donor has a central helium
mass fraction of {$Y_\mathrm{c}=0.02$}, and for the entirety of its post-CE lifetime
until core carbon depletion (26000 years), it is a Roche-lobe filling binary.
This matches the results of \citet{Quast+2019} who argue that post-CE systems
can undergo MT on the nuclear timescale of the donor. We expect that
in cases where the star does not halt its expansion in the HG, CE evolution is
initiated with a much larger value of $Y_\mathrm{c}$, leading to a longer lifetime
as an X-ray source post-CE.

\begin{figure}
   \includegraphics[width=\columnwidth]{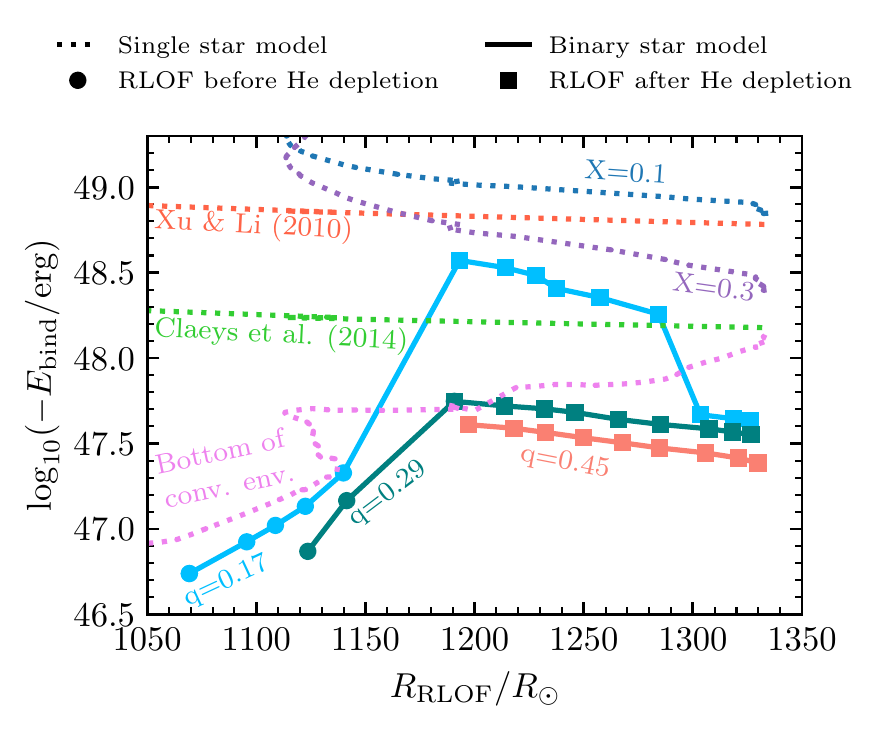}
   \caption{Binding energy of the envelope of a star with initial
   mass of $30M_\odot$ as a function of radius at
   RLOF. Solid lines indicate the results from our $\alpha_\mathrm{CE}=1$ grid with
   three different mass ratios $q=0.17$, $0.29$ and $0.45$. Dotted lines show
   the binding energy at a given radius deduced from a single star model, including
   the results obtained with the fit of \citet{XuLi2010}, the fit of
   \citet{Claeys+2014}, and assuming the bottom of the envelope to be at
   $X=0.1$, $X=0.3$ or to correspond to the bottom of the convective envelope.
   Symbols indicate whether the binary undergoes RLOF before or after core
   helium depletion.}\label{fig:Ebind_conv}
\end{figure}

The single model we considered with different values of $\alpha_\mathrm{CE}$
illustrates how the binding energy is dependent on the efficiency of CE
evolution, but using the models from our grid we can see how the resulting
binding energy of systems surviving CE evolution depends on mass ratio as well.
Fig.~\ref{fig:Ebind_conv} illustrates the binding energy as a function of radius
at RLOF resulting from models at three different initial mass ratios,
$\alpha_\mathrm{CE}=1$ and varying initial orbital periods. For most of these
simulations we find that CE evolution terminates near or above the mass
coordinate of the bottom of the convective envelope at RLOF, resulting in
binding energies $<10^{48}~\mathrm{erg}$. These low binding energies are below those
predicted by the fits of \citet{XuLi2010} and \citet{Claeys+2014}, although for
a mass ratio of $q=0.17$ we find binary models with binding energies between the
value of these two fits. The $q=0.17$ simulations with binding energies
$>10^{48}~\mathrm{erg}$ behave similarly to the simulation with $\alpha_\mathrm{CE}=0.1$
shown in Fig.~\ref{fig:ebindce}, where the binary only detaches after removing
layers with a high binding energy that were radiative at the onset of CE
evolution. This results in an order of magnitude variation of the binding energy
with varying mass ratio.

\subsection{Common envelope in donors with radiative envelopes}

As all of our simulations where CE evolution is triggered for a star with a radiative
envelope are expected to merge during CE evolution, we cannot properly define a core--envelope
boundary to compute the binding energy. However, we can
artificially produce systems that eject their outer envelope by using
artificially high values of $\alpha_\mathrm{CE}$. We consider here systems
consisting of a $30M_\odot$ donor and a $3M_\odot$ BH with initial orbital
periods between {$10^{1}$~days} and $10^{3}$~days, and values of $\alpha_\mathrm{CE}$
of $10$ and $20$. The low mass of the BH, which would fall in the putative lower BH mass gap \citep{Ozel+2010,Farr+2011,Kreidberg+2012,GW190814},
is chosen such that MT is unstable across the range of chosen
orbital periods.

\begin{figure}
   \includegraphics[width=\columnwidth]{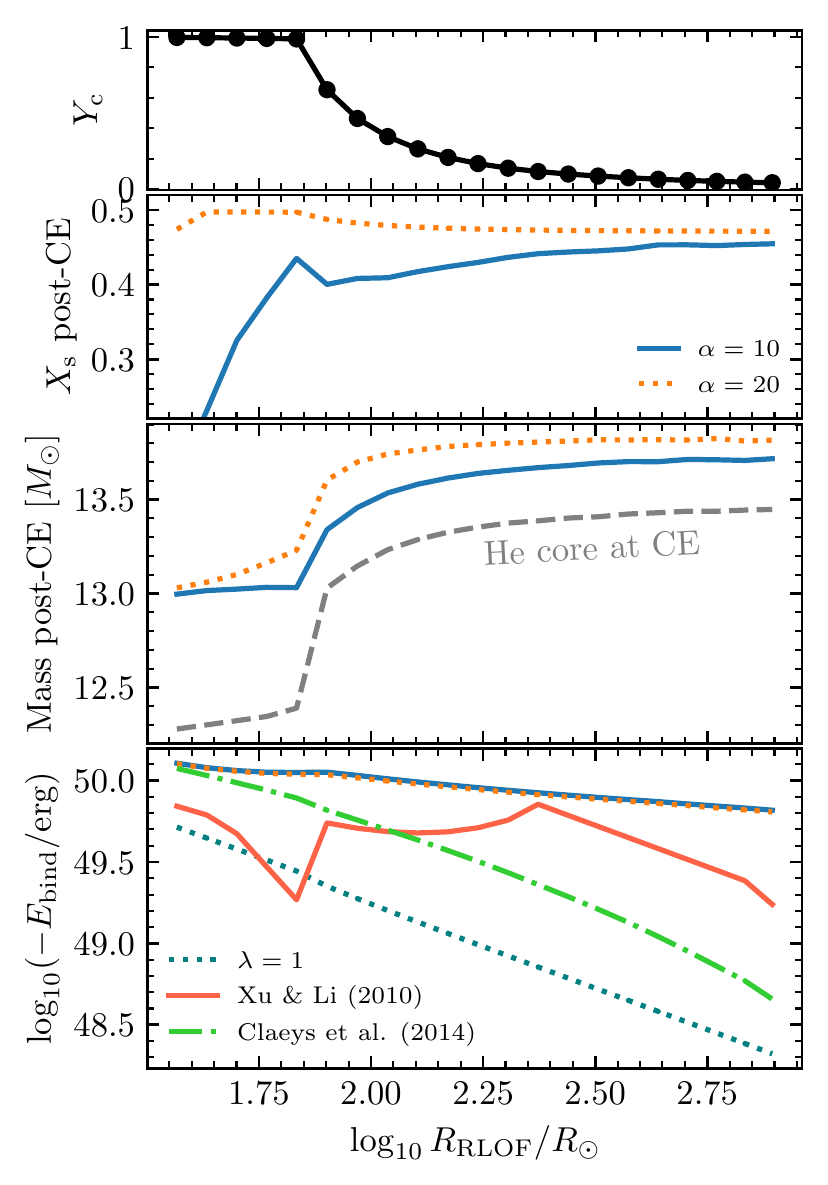}
   \caption{Properties after CE evolution for the case of donor stars with a
   radiative envelope, with artificially high values of $\alpha_\mathrm{CE}=10,20$
   in order to allow for a successful CE ejection. Simulations
   correspond to binary systems composed of a star with a ZAMS mass of
   $30M_\odot$ and a BH companion of $3M_\odot$, with periods ranging from
   {$10^{1}$~days} to $10^{3}$~days. Results are shown as a function of the radius of
   the star when it fills its Roche lobe. \textit{Top panel}: Central helium mass fraction
   $Y_\mathrm{c}$ at
   the onset of CE, showing interaction before and after the end of the HG as a
   large drop in $Y_\mathrm{c}$. \textit{Second panel}: Surface hydrogen mass fraction
   after envelope ejection. \textit{Third panel}: Stellar mass after envelope ejection,
   showing also the mass of the helium core defined as the innermost mass
   coordinate with $Y>0.01$. \textit{Bottom panel}: Binding energy of the ejected
   envelope, as well as the binding energy that would be predicted by the
   \citet{Claeys+2014} and \citet{XuLi2010} fits, and the one corresponding to
   $\lambda=1$.}\label{fig:cerad}
\end{figure}

All of these simulations manage to eject their envelopes during
CE evolution, and Fig.~\ref{fig:cerad} shows the resulting masses post-CE as
well as the binding energy of the ejected layers. We find that CE terminates
when the surface hydrogen mass fraction is relatively high
{($X\sim0.4$--$0.5$)}, and
that the binding energy of the envelope {varies between $1.3\times
10^{50}$~erg and $6.6\times 10^{49}$~erg, a factor of $2$ variation despite a
change by a factor of $20$ in radius at RLOF}. Figure~\ref{fig:cerad} also illustrates that
interaction happening before or after the end of the HG has no impact on the
binding energy. Estimating the binding energy using either the
\citet{Claeys+2014} or \citet{XuLi2010} fits
would underestimate the binding energy of the envelope for all these simulations.
Since the binding energy {just changes by a factor of $2$} despite
{over an} order of magnitude change
in the radius of the star, population synthesis calculations that approximate
$E_\mathrm{bind}$ using a fixed value of the dimensionless $\lambda$ parameter
cannot reproduce the variation of $E_\mathrm{bind}$ with radius, independent of the
choice of $\lambda$.

\section{Discussion} \label{sec:discussion}

We have studied the outcomes from interaction of
low-metallicity massive stars with a companion BH using detailed binary
evolution calculations performed with the \texttt{MESA} code. Our simulations,
which consist of a $30M_\odot$ low-metallicity donor with a companion BH, show several features
relevant to understand the formation of merging binary BHs through both
stable MT and CE evolution. For these simulations we have also updated the MT
prescription developed by \citetalias{KolbRitter1990}, including the possibility of
overflow from an outer Lagrangian point, and we also implemented a method to
self consistently determine the core--envelope boundary in cases of CE evolution.
The main conclusions we derive from our simulations are:
\begin{enumerate}
   \item Stable MT after the formation of the first BH in a binary can produce merging binary BHs for a broad range
      of orbital periods. The boundary between stable and unstable MT
      is located at a point that allows for orbital hardening just from stable
      MT, without requiring a CE phase \citep{vandenHeuvel+2017}. This includes
      cases where MT between the star and the first formed BH is initiated
      during the main sequence. We also find cases where the low density
      envelope of the donor would overflow its outer Lagrangian point. Using our
      model to quantify this outflow, we find that binaries can remain stable
      despite the high specific angular momentum of this outflow, 
      resulting in more compact binaries after MT than if the process is ignored.
   \item We do not find any case of CE evolution initiated with a star with a
      radiative envelope where the envelope is successfully ejected. Between the
      TAMS and the start of the rise in the Hayashi line, the binding energy of
      the envelope remains at $\sim10^{50}~\mathrm{erg}$, {varying by just a
      factor of two} despite an increase of more
      than an order of magnitude in radius. Before the formation of a deep
      convective envelope, we find that the prescriptions of \citet{XuLi2010},
      {as developed by \citet{Dominik+2012} for $M>20M_\odot$,} and
      \citet{Claeys+2014} underestimate the binding energy of the envelope by
      {up to an} order of magnitude.
    \item The parameter space where systems
      can survive CE is limited to a small range in orbital period ($\sim 0.2~\mathrm{dex}$) where RLOF happens after the formation of a deep convective envelope.
      Assuming that all internal and recombination energy as well
      as all the energy from inspiral during CE is used to eject the envelope, we find the majority of these
      systems surviving CE would form binary BHs too wide to merge within a
      Hubble time.
   \item As in \citet{DeloyeTaam2010}, we find that the core--envelope boundary is not uniquely
      defined, with different CE efficiencies (or similarly,
      different mass ratios) resulting in different binding energies for the
      envelope. In particular for the case discussed in Sec.~\ref{sec:CE}, we
      compute binding energies that differ by over an order of magnitude
      depending on the efficiency of CE evolution. This is in contrast to the
      way binding energies are determined in population synthesis codes,
      where a star has a uniquely defined binding energy at each point of its
      evolution.
   \item Under the simple assumption of a flat distribution in mass ratio and a
      flat distribution in $\log P$ after the formation of the first BH, we find
      that stable MT would dominate the formation rate of merging binary BHs.
      If we take CE efficiency to be $100\%$, the ratio of binaries surviving CE evolution (independent of
      whether they form a compact or wide binary BH) to those that form merging
      binary BHs from stable MT, is {$0.35$}. This is further reduced
      to {$0.017$} if we
      take only the CE simulations that produce merging binary BHs. This is in
      significant contradiction with rapid population synthesis calculations
      that predict the vast majority ($>90\%$) of BHs are formed via 
      an initial phase of stable MT before the formation of the first
      BH, followed by a CE phase when the secondary fills its Roche lobe (cf.\
      Table~4 of \citealt{Dominik+2012}).
\end{enumerate}
These results broadly agree with those of \citet{Klencki+2020a} and
\citet{Klencki+2020b}, who have used single star models with a fixed definition
of the core--envelope boundary to propose current population synthesis
calculations may overestimate the production rate of merging binary BHs from CE
evolution. Since our simulations are restricted to a $30M_\odot$ donor at
$Z=Z_\odot/10$, it is necessary to further explore different masses and
metallicities to understand how generally our conclusions apply (Gallegos-Garcia
et al., in preparation).
{
Recent investigations using rapid population synthesis codes have indicated that
given current uncertainties in binary evolution such as the cosmic star
formation rate \citep{Neijssel+2019} or the critical mass ratio for stable MT
\citep{Andrews+2020}, stable MT could be an important production channel for
the merging binary BHs observed with gravitational waves.
Our results indicate these rapid population synthesis
calculations still significantly overestimate
the number of systems surviving CE evolution at low metallicities, and
accounting for this would further reduce the contribution from the CE evolution
channel.
}

As we find that the boundary between stable and unstable MT
is located right at a point that allows for the formation of merging
binary BHs via stable MT, any uncertainties that can shift this boundary can
have a large effect on the number and properties of expected GW sources. Among
such uncertainties, we can point out that our simulations do not include stellar
rotation or tidal coupling to the orbit. This additional sink of orbital angular
momentum can potentially affect the stability of MT in our calculations.
We have explicitly checked in our stable MT simulations that the moment
of inertia of the star is too low to undergo the Darwin instability
\citep{Darwin1879}, which is the limit were tidal coupling leads to a runaway
shrinkage of the orbit. Another important simplification is that we do not account
for the effect of the companion in the momentum equation, which would make the
adiabatic response of the star to mass loss dependent on the mass ratio of the
system and its separation. It has also been argued that hydrodynamics need to be
considered to assess the stability of MT for giant donors
\citep{PavlovskiiIvanova2015}, although a consistent treatment of hydrodynamics
needs to consider as well the perturbation to the gravitational potential from
the companion star. {Additionally, a significant amount of the systems we predict that form
merging binary BHs also undergo overflow of the L$_2$ and even the L$_3$
equipotentials. Although for these large overflow cases we have verified merging
binary BHs are formed even under extreme assumptions of mass and angular
momentum outflows, three-dimensional hydrodynamic studies are required to
accurately model outflows in such systems}. All these uncertainties need to be studied in detail to
determine if stable MT can indeed produce merging binary BHs.

Our simulations for the outcome of CE evolution also have important
uncertainties. Although we do not have the core--envelope mass boundary as a free parameter
we still have the CE efficiency as an unknown variable.
The efficiency can be reduced by radiative processes or also by a surplus of
energy being removed as kinetic energy from the system \citep[cf.][]{NandezIvanova2016}.
Within the context of one-dimensional stellar models, it might be possible to constrain the
efficiency of the CE process by means of simplified hydrodynamical simulations
of the inspiral \citep{Taam+1978,Clayton+2017,Fragos+2019} or, ideally, with
full three-dimensional simulations of CE evolution in massive stars
\citep{TaamSandquist2000,Ricker+2019,Law-Smith+2020}. Despite these uncertainties that
significantly affect the outcome of systems that survive CE evolution, we consider
our conclusion that massive stars would merge when undergoing CE prior to the
formation of a convective envelope as much more certain. We find this result for
the case of full efficiency for the CE process, including recombination energy.
Our method to determine the core--envelope boundary can also be considered to
produce a lower bound on the binding energies. If we were to consider the
boundary deeper inside down at the point where thermal pulses would not happen
post-CE \citep{Ivanova2011} we would find larger binding energies and less
systems surviving CE evolution.

\begin{figure}
   \includegraphics[width=\columnwidth]{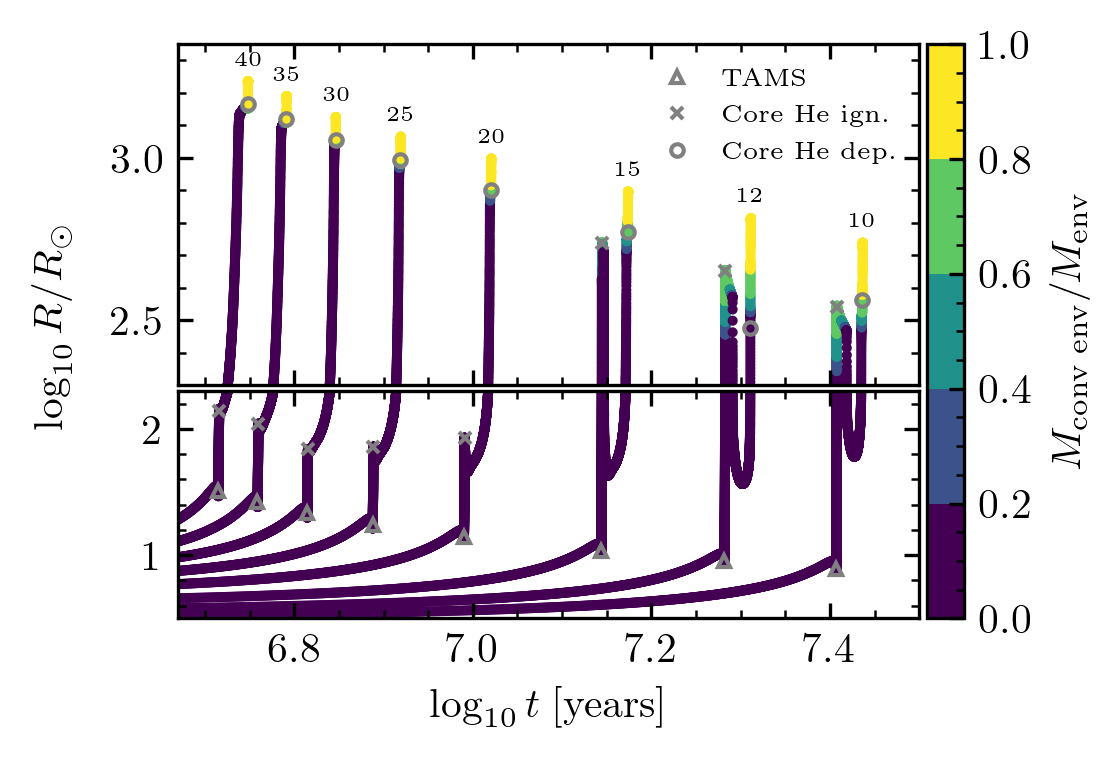}
   \caption{Radial evolution of single stars of various masses at a metallicity
   of $Z_\odot/10$ modeled until core-carbon depletion. Numbers above each track indicate the mass of the model in
   solar masses, and symbols indicate different stages in the evolution of the
   star. Color of the tracks indicates the fraction of the hydrogen envelope
   that is part of an outer convective zone.}\label{fig:radius}
\end{figure}

Our simulations do not exclude CE evolution
as an efficient mechanism to form merging binary neutron stars. At lower masses,
a larger fraction of the stellar expansion of a star during its lifetime happens after the
formation of a deep convective envelope (Fig.~\ref{fig:radius}). This results in a broader range of
orbital periods where we expect the ejection of a CE to be
possible, leading to binaries compact enough to merge from GW radiation.
Following the work of \citet{Misra+2020}, the boundaries for stable MT
between a star and a neutron star companion do not allow for stars massive enough
to form a second neutron star in the system. This excludes stable MT as a
formation channel for merging NSs. The case of
MT and CE evolution in the case of neutron star companions will be studied in future
work.

\begin{acknowledgements}
{We thank the referee for detailed feedback on our work.} 
PM acknowledges support from the FWO junior postdoctoral fellowship No.\
   12ZY520N. 
M.G-G. is grateful for the support from the Ford Foundation Predoctoral
   Fellowship.
CPLB is supported by the CIERA Board of Visitors Professorship.
VK is supported by a Senior CIERA Fellowship and by Northwestern University. PM
thanks Alina Istrate for useful discussion.
\end{acknowledgements}
\bibliographystyle{aa}

\appendix
\section{Computation of $\mathrm{d} A/\mathrm{d}\Phi$}\label{app:dadphi}

The coefficients of the quadratic terms in the Taylor expansion of Eq.~\eqref{equ:taylor} are given by
\begin{eqnarray}
\begin{split}
   C_1=&\frac{1}{2}\left(\frac{\partial^2 \Phi}{\partial y^2}\right)_{{\rm
      L}1}\nonumber\\
      =&\frac{G M_d}{2a^3}\left[\frac{1}{|\hat{X}_{\rm
      L1}|^3}+\frac{q}{|1-\hat{X}_\mathrm{L1}|^3}-(q+1)\right],\nonumber
   \\
      C_2=&\frac{1}{2}\left(\frac{\partial^2 \Phi}{\partial z^2}\right)_{{\rm
      L}1}\nonumber\\
    =&\frac{G M_d}{2a^3}\left[\frac{1}{|\hat{X}_{\rm
      L1}|^3}+\frac{q}{|1-\hat{X}_\mathrm{L1}|^3}\right],
\end{split}
\end{eqnarray}
and the coefficients of the quartic terms are
\begin{eqnarray}
\begin{split}
   C_3=&\frac{1}{24}\left(\frac{\partial^4 \Phi}{\partial y^4}\right)_{{\rm
      L}1}\nonumber\\
      =&-\frac{3G
      M_d}{8 a^5}\left[\frac{1}{|\hat{X}_\mathrm{L1}|^5}+\frac{q}{|1-\hat{X}_{\rm
      L1}|^5}\right],\nonumber
   \\
      C_4=&\frac{1}{4}\left(\frac{\partial^4 \Phi}{\partial y^2 z^2}\right)_{{\rm
      L}1}=2C_3\nonumber
   \\
      C_5=&\frac{1}{24}\left(\frac{\partial^4 \Phi}{\partial z^4}\right)_{{\rm
      L}1}=C_3.
\end{split}
\end{eqnarray}
When we consider outflows from the outer Lagrangian point of the donor,
$\hat{X}_{{\rm L}1}$ is switched by the normalized $x$-coordinate of the outer Lagrangian
point, $\hat{X}_\mathrm{L\; out}$, which with our definition has a negative value.

In order to estimate the area on the plane crossing L$_1$ for a given
$\Delta\Phi$, it is useful to use cylindrical coordinates $y=\varpi \sin\theta$,
$z=\varpi \cos\theta$, in which case
\begin{eqnarray}
   \varpi^2=\frac{1}{\gamma}\Delta\Phi - \frac{C_3}{\gamma^3}(\Delta\Phi)^2
   +\mathcal{O}\left[(\Delta \Phi)^3\right],
\end{eqnarray}
where $\gamma \equiv C_1\sin^2\theta + C_2\cos^2\theta$. This can be integrated
to obtain the area $A(\Phi)$,
\begin{eqnarray}
   A=2\int_0^{\pi/2} \varpi^2 \,\mathrm{d}\theta,
\end{eqnarray}
which after differentiation by $\Phi$ gives Eq.~\eqref{equ:dadphi_f}.

\begin{figure}
   \includegraphics[width=\columnwidth]{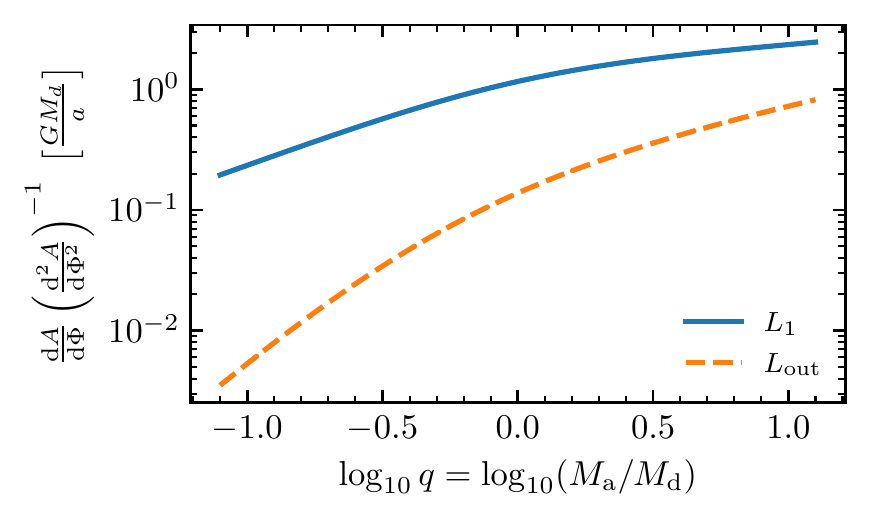}
   \caption{Ratio of $\mathrm{d}A/\mathrm{d}\Phi$ to to
   $\mathrm{d}^2A/\mathrm{d}\Phi^2$ at L$_1$ and at the outer
   Lagrangian point of a star expressed in units of $GM_\mathrm{d}/a$.
   }\label{fig:dadphi}
\end{figure}

The relative importance of the constant and linear terms for $\mathrm{d}A/\mathrm{d}\Phi$ shown in
Eq.~\eqref{equ:dadphi_f} can be assessed by taking the ratio of $\mathrm{d}A/\mathrm{d}\Phi(\Delta
\Phi=0)$ and $\mathrm{d}^2A/\mathrm{d}\Phi^2(\Delta \Phi=0)$. This essentially quantifies how
large $\Delta \Phi$ needs to be to produce a significant change in $\mathrm{d}A/\mathrm{d}\Phi$,
with smaller values indicating a higher sensitivity to $\Delta \Phi$, and thus
a larger importance of the higher-order term.
Figure~\ref{fig:dadphi} shows this value at different mass ratios both for the
L$_1$ and L$_\mathrm{out}$ points of a star. This value is always positive,
indicating that the inclusion of the higher-order term results in an increased
value of the computed MT rate. We also see that the value of the ratio is
much smaller for the outer Lagrangian point, and in the case where the donor is
the more massive component in the binary, a variation of $\Phi$ on the order of
$0.1 GM_\mathrm{d}/a$ can modify $\mathrm{d}A/\mathrm{d}\Phi$ significantly. This points out that the
inclusion of the higher-order term has a larger impact on the computation of
outflows from the outer Lagrangian point than from L$_1$.

\section{Dependence on numerical parameters of CE algorithm}\label{app:numerics}
\begin{figure}
   \includegraphics[width=\columnwidth]{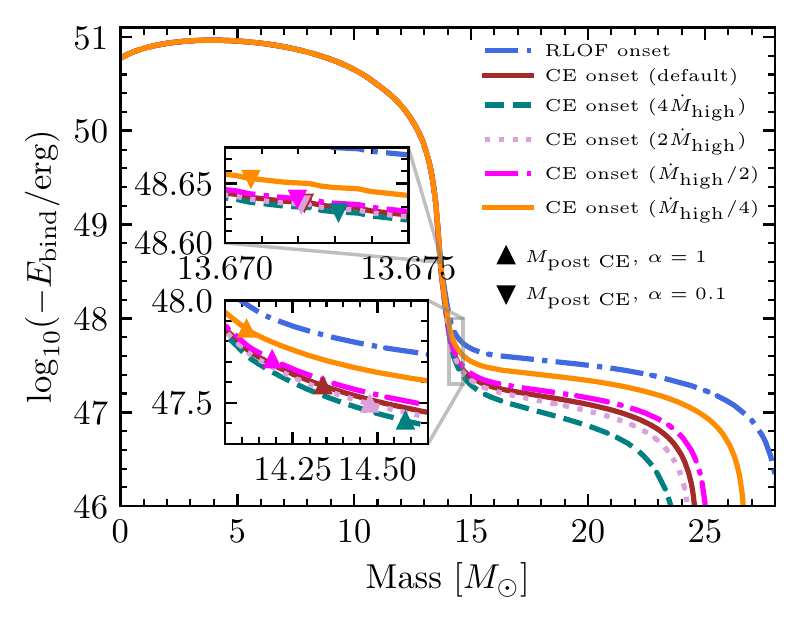}
   \caption{Binding energy profiles of the donor in a system with an initial mass of
   $30M_\odot$ and a BH companion of {$14.1M_\odot$} at an orbital period of
   {$2344$~days}. Profiles are shown at RLOF and at the onset of CE evolution for
   different values of $\dot{M}_\mathrm{high}$. Symbols indicate the mass
   coordinate at which CE evolution terminates for each choice of $\dot{M}_{\rm
   high}$ and for CE efficiencies of $\alpha_\mathrm{CE}=1$ and $\alpha_\mathrm{CE}=0.1$.}\label{fig:ebindce2}
\end{figure}

As our algorithm to model CE evolution has various numerical parameters, it is
important to evaluate in detail how our results depend upon these. Here we
modify the three parameters described in Sec.~\ref{sec:CE}: $\dot{M}_{\rm
high}$, $\dot{M}_\mathrm{low}$ and $\delta$, for which in our default model we use
$1\;M_\odot\;{\rm yr^{-1}}$, $10^{-5}\;M_\odot\;{\rm yr^{-1}}$ and $0.02$
respectively. We consider here the difference in the outcome of the CE simulations
shown in Sec.~\ref{sec:ce_examples} for variations in these three parameters,
with the properties of the system post-CE being shown in Table \ref{table:CE}.
For all parameters we find only $\sim 5\%$ variations in the post-CE properties,
except for variations of $\dot{M}_\mathrm{high}$ when using an efficiency $\alpha_{\rm
CE}=1$. Detachment in these systems happens in the proximity of the mass
coordinate of the bottom of the convective envelope at the onset of CE.
Variations in $\dot{M}_\mathrm{high}$ modifies the mass of the star at the onset of
CE, as is shown in Table~\ref{table:CE}. This modifies the binding energy
in the outermost layers, which is illustrated in Fig.~\ref{fig:ebindce2}, making evolution sensitive to the exact point at which
we switch between MT and CE evolution. This is not the case for the system with
$\alpha_\mathrm{CE}=0.1$, as the core--envelope boundary is computed to be deeper
in, in layers that are not affected by the exact point at which we switch
between MT and CE evolution.

\begin{table*}
\caption{Variation in the outcome of our CE simulations with respect to the
   numerical parameters of the model. Results are shown for the system with a
   {$14.1M_\odot$} BH and an initial period of
   {$2344$~days} which is described in
   Section \ref{sec:ce_examples}. The default choices for the numerical
   parameters are $\dot{M}_\mathrm{high}=1\;M_\odot\;{\rm yr^{-1}}$, $\dot{M}_{\rm
   low}=10^{-5}\;M_\odot\;{\rm yr^{-1}}$ and $\delta=0.02$. The variations of
   these parameters that we consider correspond to varying $\dot{M}_\mathrm{high}$
   and $\dot{M}_\mathrm{low}$ by up to a factor of $4$, and modifying $\delta$ to
   $0.001$ and $0.04$. For each case the first two columns show the mass and
   separation of the system at the onset of CE, while the rest of the columns
   show the resulting core mass $M_\mathrm{core}$, binding energy $E_\mathrm{bind}$
   and post-CE separation $a_\mathrm{post\;CE}$ derived by the simulations for both
   $\alpha_\mathrm{CE}=1$ and $\alpha_\mathrm{CE}=0.1$.
    }              
\label{table:CE}      
\centering                                      
\begin{tabular}{c cc@{\extracolsep{4pt}}  ccc@{\extracolsep{4pt}}  ccc}          
\hline\hline                        
   & &  &  & $\alpha_\mathrm{CE}=1$ &  & & $\alpha_\mathrm{CE}=0.1$ & \\    
   \cline{4-6} \cline{7-9}
   Model & $M_\mathrm{pre\,CE}\,{\rm [M_\odot]}$ & $a_\mathrm{pre\,CE}\,{\rm [R_\odot]}$ &
   $M_\mathrm{core}\,{\rm [M_\odot]}$ & $-E_\mathrm{bind}\,{\rm [erg]}$ & $a_{\rm
   post\,CE}\,{\rm [R_\odot]}$ & $M_\mathrm{core}\,{\rm [M_\odot]}$ & $-E_{\rm
   bind}\,{\rm [erg]}$ & $a_\mathrm{post\,CE}\,{\rm [R_\odot]}$ \\    
\hline                                   
   default & 24.71 & 2335 &  14.34 & $3.876\times 10^{47}$ & 572.5 &  13.67 &
   $4.327\times
   10^{48}$ & 8.414\\
\hline                                   
   $4\dot{M}_\mathrm{high}$ & 23.89 & 2215 &  14.58 & $2.619\times10^{47}$ &
   710.2 & 13.67
   & $4.254\times10^{48}$ & 8.556\\
   $2\dot{M}_\mathrm{high}$ & 24.44 & 2307 & 14.48 & $3.160\times 10^{47}$ &
   646.7 & 13.67 & $4.314\times 10^{48}$ & 8.439 \\
   $\dot{M}_\mathrm{high}/2$ & 25.15 & 2372 & 14.19 & $5.192\times 10^{47}$ & 473.3
   & 13.67 & $4.363\times 10^{48}$ & 8.343 \\
   $\dot{M}_\mathrm{high}/4$  & 26.77 & 2529 & 14.11 & $7.280\times 10^{47}$ &
   373.9 & 13.67 & $4.539\times 10^{48}$ & 8.021 \\
\hline                                   
   $4\dot{M}_\mathrm{low}$ & 24.71 & 2335 & 14.34 & $3.876\times 10^{47}$ & 572.5 &
   13.68 & $3.979\times 10^{48}$ & 9.231\\
   $\dot{M}_\mathrm{low}/4$ & 24.71 & 2335 & 14.34 & $3.876\times 10^{47}$ & 572.5
   & 13.67& $4.314\times 10^{48}$ & 8.436 \\
\hline                                   
   $\delta\hspace{-0.03in}=\hspace{-0.03in}0.001$ & 24.71 & 2335 & 14.34 &
   $3.876\times 10^{47}$ & 571.1 & 13.67 & $4.267\times 10^{48}$ & 8.531 \\
   $\delta\hspace{-0.03in}=\hspace{-0.03in}0.04$ & 24.71 & 2335 & 14.34 &
   $3.876\times 10^{47}$ &
   572.5 & 13.67 & $4.385\times 10^{48}$ & 8.302 \\
\hline                                             
\end{tabular}
\end{table*}
{
\section{Comparison to other stellar evolution tracks}\label{app:codes}
A comparison of our single metal-poor $30M_\odot$ model to similar stellar evolution
tracks that are publicly available is shown in Fig.~\ref{fig:codes}. These
include tracks by \citet{Meynet+1994}, \citet{Pols+1998}, \citet{Tang+2014} and
\citet{Klencki+2020a}. Out of these the only ones that use the same stellar
evolution code are our own models and those of \citet{Klencki+2020a}. From the models of
\citet{Meynet+1994} we took the $25M_\odot$ stellar track which is the closest
one to $30M_\odot$. The track taken from \citet{Pols+1998} corresponds to their
models with overshooting, and the one from \citet{Klencki+2020a} is from their
set of non-rotating stellar tracks. All tracks experience a long-lived
phase as blue supergiants, halting their rapid expansion in the HG phase at
$\log_{10}(T_\mathrm{eff}/\mathrm{K})\sim 4.2$ (except the lower mass one from
\citealt{Meynet+1994} that does so at a higher $T_\mathrm{eff}$).
\begin{figure}
   \includegraphics[width=\columnwidth]{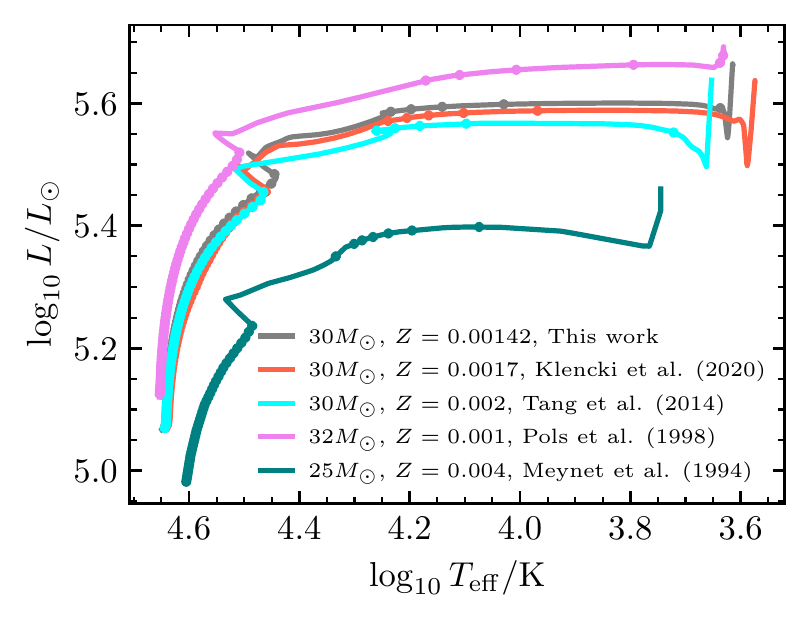}
   \caption{Evolution of stars with a mass of $\sim30M_\odot$ and a metallicity
   of $Z_\odot/10$ computed by different research groups and codes. Dots are
   plotted every $10^5$~years.}\label{fig:codes}
\end{figure}

\section{Single star resolution test}\label{app:resolution}
Given the formation of multiple intermediate convection zones in our models
during the HG phase we performed a resolution test to see if our results are
significantly modified by increased spatial and temporal resolution. We make use
of the \texttt{mesh\_delta\_coeff} and \texttt{time\_delta\_coeff} options that
allow for an approximate scaling of all spatial and temporal resolution
controls, and take models with approximately two, four, and eight times the resolution of our
default setup. The resulting tracks in the HR diagram are shown in
Fig.~\ref{fig:convergence}. We see small variations with changes in resolution
on the evolution during the HG phase producing a variation of about $10\%$ on
the effective temperature at which the HG phase completes.
\begin{figure}
   \includegraphics[width=\columnwidth]{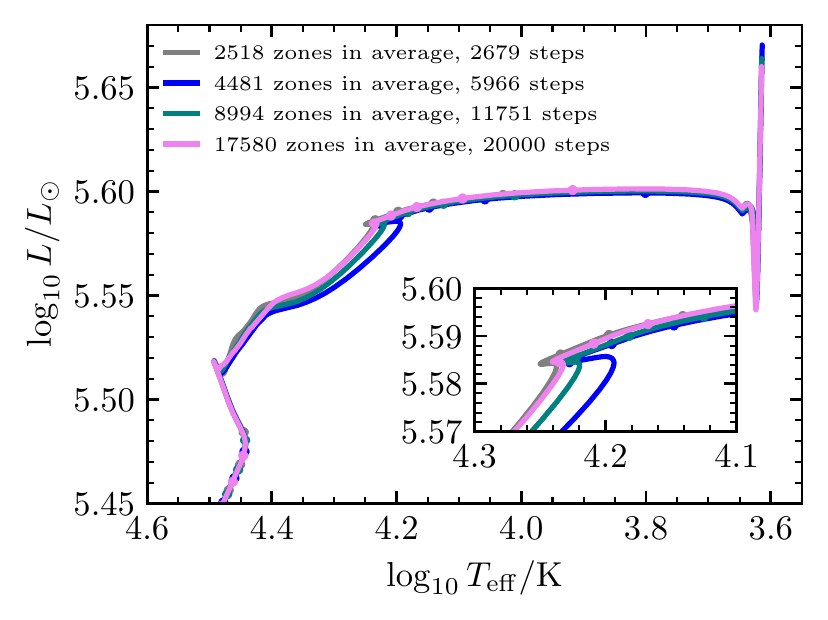}
   \caption{Evolution in the HR diagram of a single $30M_\odot$ model with a
   metallicity of $Z_\odot/10$ computed with increasing spatial and
   temporal resolution. For each track we indicate the average number of zones
   in the model as well as the steps taken in the simulation. The lowest resolution track shown corresponds to our
   default setup.}\label{fig:convergence}
\end{figure}

In particular, for this work we are interested in how the binding energy of the
envelope of a star compares to prescriptions present in the literature.
Fig.~\ref{fig:convergence_ebind} shows the binding energies obtained for models
at increasing temporal and spatial resolution, as well as the results we obtain
with the \citet{XuLi2010} and \citet{Claeys+2014} fits. Binding energies are
computed from the stellar models taking the bottom of the envelope at the
innermost mass coordinate where $X>0.1$ and $X>0.3$. Between the different
models, we see variations on the computed binding energies at a given radius
smaller than $12\%$. The larger variations happen at the end of the HG and are
due to the slightly different radii achieved at the end of the HG. Ignoring the
values between $\log_{10}(R/R_\odot)=1.8$--$1.95$ we find that the differences are
smaller than $6\%$.
\begin{figure}
   \includegraphics[width=\columnwidth]{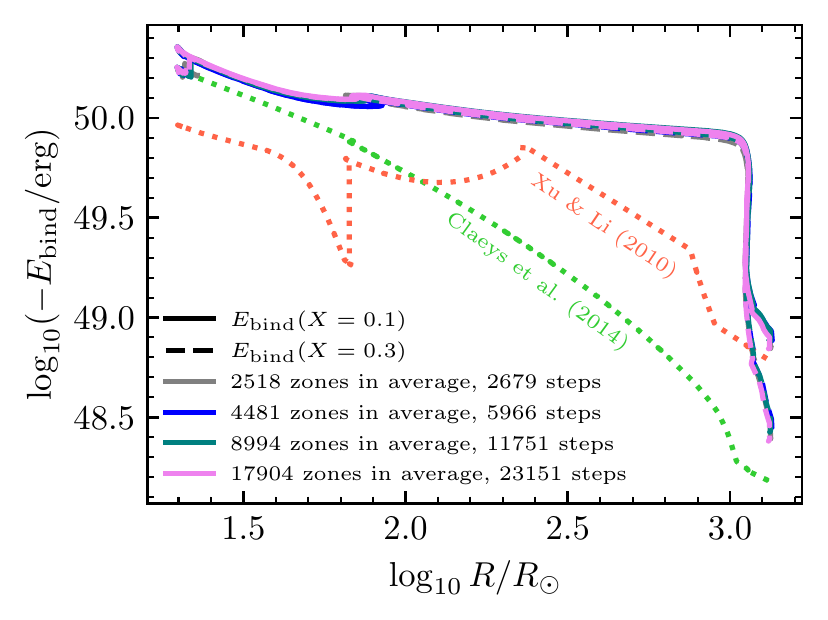}
   \caption{Binding energy of the envelope of a post-main-sequence $30M_\odot$ model with a
   metallicity of $Z_\odot/10$ computed with increasing spatial and
   temporal resolution. For each track we indicate the average number of zones
   in the model as well as the steps taken in the simulation. The lowest resolution track shown corresponds to our
   default setup. For comparison we include the binding energies obtained using
   our default resolution together with the fits of \citet{XuLi2010} and
   \citet{Claeys+2014}. Before the formation of a convective envelope
   the binding energies computed at $X=0.1$ and $X=0.3$ almost overlap.}\label{fig:convergence_ebind}
\end{figure}
\section{Application of $\lambda$ fits to stellar models}\label{app:lambda}
The fits from \citet{Claeys+2014} and \citet{XuLi2010} depend on the mass of the
star, its radius, and its evolutionary stage in terms of the star-types defined
by \citet{Hurley+2002}.
A model analogous to our $30M_\odot$ model is assigned
types "HG" during the HG, "CHeB" during core helium burning but before ascending
the Hayashi line and "EAGB" while on the Hayashi line. To distinguish between
the CHeB and EAGB types in our models we take models with convective envelopes
$>1M_\odot$ to be of type EAGB, as we find this is close to the point where the
star rises on the Hayashi line. As we are only concerned with these three
assigned types, we only describe the fits associated to them.
\subsection{\citet{Claeys+2014}}
The fits of \citet{Claeys+2014} were first presented by \citet{Izzard2004}, who
took them directly from the source code of the \texttt{BSE} code
\citep{Hurley+2002}.
\citet{Claeys+2014} do not specify the range
of masses used to compute these fits.
The fit for $\lambda$ including internal but not
recombination energy is
\begin{eqnarray}
   \lambda=2\times
   \begin{cases}
      \lambda_2 & M_\mathrm{conv\;env}=0\\
      \lambda_2 + M_\mathrm{conv\;env}^{0.5}(\lambda_1-\lambda_2) & 0<M_{\rm
      conv\;env}=M_\odot\\
      \lambda_1 & M_\mathrm{conv\;env}>M_\odot
   \end{cases},
\end{eqnarray}
where
\begin{eqnarray}
   \lambda_2=0.42\left(\frac{R_\mathrm{ZAMS}}{R}\right)^{0.4},
\end{eqnarray}
with $R_\mathrm{ZAMS}$ being the radius of the stellar model at the ZAMS. The value
of $\lambda_1$ is given by
\begin{eqnarray}
   \lambda_1 = \min\{0.8,1.25-0.15\log_{10} (L/L_\odot),\lambda_3\}
\end{eqnarray}
with
\begin{eqnarray}
   \lambda_3 = \min\{0.9,0.58+0.75\log_{10}(M/M_\odot)\}-0.08\log_{10}(L/L_\odot).
\end{eqnarray}
This fixes a typo ($0.9$ instead of $-0.9$ in the minimum function)
from \citet{Claeys+2014} and \citet{Izzard2004} that
is inconsistent with the implementation in the \texttt{BSE} code from which the
fit was taken. \citet{Claeys+2014} also describes variations to the $\lambda_1$ and
$\lambda_3$ values for stellar types that are not relevant to our $30M_\odot$
model, so we do not describe those.
\subsection{\citet{XuLi2010}} \label{subapp:xuli}

\begin{table*}
   \caption{Fitting coefficients for the $\lambda$ fits of \citet{XuLi2010}
   including thermal energy, as
   implemented by \citet{Dominik+2012} for Population II stars with masses
   $18M_\odot<M<35M_\odot$. Values are taken from the source code of the
   \texttt{COMPAS} code. Exact numbers from the source code are included here,
   so the number of significant digits is variable.}              
\label{table:nanjing}      
\centering                                      
\begin{tabular}{ccccc}          
\hline\hline                        
   Star type & $a_1$ & $a_2$ & $a_3$ & $a_4$ \\
\hline                                   
   HG        & $1.27138$ & $0.00538$ & $-0.0012$ & $1.80776\times 10^{-5}$\\
   CHeB        & $0.69746$ & $-0.0043$ & $8.97312\times10^{-6}$ & $-5.83402\times 10^{-9}$ \\
   EAGB        & $-436.00777$ & $1.41375$ & $-0.00153$ & $5.47573\times 10^{-7}$\\
\hline                                             
\end{tabular}
\end{table*}
The $\lambda$ fits of \citet{XuLi2010} cover up to a mass of $20M_\odot$, while
fits for higher masses where developed by \citet{Dominik+2012}. These fits are
not published, but can be found in the source code of the \texttt{COMPAS} code.
Fits are provided for the binding energy with and without thermal energy
included, i.e.\ $\alpha_\mathrm{th}=0$ or $1$ in Eq.~\eqref{equ:ebindint}. As we
work with $\alpha_\mathrm{th}=1$ we compare to the $\lambda$ fit including thermal
energy. Fits are provided individually for each of the HG, CHeB and EAGB phases,
and the ones that correspond to our $30M_\odot$ model at $Z_\odot/10$ are the
Population II fits for masses $18M_\odot<M<35M_\odot$. These are:
\begin{itemize}
   \item HG:
      \begin{eqnarray}
         \lambda = 
         \begin{cases}
            a_1+a_2\xi+a_3\xi^2+a_4\xi^3 & R\le 900R_\odot\\
            0.2 & R>900R_\odot,
         \end{cases}
      \end{eqnarray}
      where $\xi=R/R_\odot$ and the value of the fitting coefficients is given in
      Table~\ref{table:nanjing}.
   \item CHeB:
      \begin{eqnarray}
         \lambda = 
         \begin{cases}
            a_1+a_2\xi+a_3\xi^2+a_4\xi^3 & R\le 230R_\odot\\
            0.1 & 230R_\odot<R<755R_\odot\\
            0.1b + 0.2 (1-b) & 755R_\odot<R<900R_\odot\\
            0.2 & R>900R_\odot,
         \end{cases}\label{equ:CHeB}
      \end{eqnarray}
      with the expression for the range $755R_\odot<R<900R_\odot$ being a linear
      interpolation from $0.1$ to $0.2$ with $b=(900-\xi)/(900-755)$.
   \item EAGB:
      \begin{eqnarray}
         \lambda = 
         \begin{cases}
            a_1+a_2\xi+a_3\xi^2+a_4\xi^3 & R\le 725R_\odot\\
            0.1 & 725R_\odot<R<850R_\odot\\
            0.1b + 0.2 (1-b) & 850R_\odot<R<900R_\odot\\
            0.2 & R>900R_\odot,
         \end{cases}
      \end{eqnarray}
      where $b=(900-\xi)/(900-850)$.
\end{itemize}
All resulting values are capped between $0.05$ and $1.5$. The
interpolation in the range $755R_\odot$--$900R_\odot$ for CHeB stars is missing in
the implementation of \citet{Dominik+2012}, and it erroneously applies the
fitting polynomial in this range resulting in negative values, producing a
positive value only because the final result is capped to a minimum of $0.05$.
Not including this correction and having $\lambda=0.05$ in this range of radii
does not change our conclusions, as the binding energy is still underestimated
compared to our own models.
}
\end{document}